\documentclass[nofootinbib, amsmath,amssymb, aps,]{revtex4}

\usepackage{hyperref}

\usepackage{graphicx}
\usepackage{dcolumn}
\usepackage{bm}

\usepackage{booktabs}
\usepackage{topcapt}

\usepackage{xcolor}

\usepackage{slashed} 

\usepackage{axodraw2}

\usepackage[T1]{fontenc}
\usepackage{color}   
\usepackage{hyperref}
\hypersetup{
    colorlinks=true, 
    linktoc=all,     
    linkcolor=blue,  
}

\newcommand{\be}{\begin{equation}}
\newcommand{\ee}{\end{equation}}
\newcommand{\bea}{\begin{eqnarray}}
\newcommand{\eea}{\end{eqnarray}}

\newcommand{\bit}{\begin{itemize}}
\newcommand{\eit}{\end{itemize}}

\begin{document}

\title{ Four Gluon Vertex from Lattice QCD}

\author{Manuel Cola\c{c}o}
\email{manuel.sc.colaco@gmail.com}
\author{Orlando Oliveira}
\email{orlando@uc.pt}
\author{Paulo J. Silva}
\email{psilva@uc.pt}
\affiliation{CFisUC, Departamento de F\'{\i}sica, Faculdade de Ci\^encias e Tecnologia, Universidade de Coimbra, 3004-516 Coimbra, Portugal}

\begin{abstract}
A lattice QCD calculation for the four gluon one-particle irreducible Green function in the Landau gauge is discussed. Results for
some of the associated form factors are reported for kinematical configurations with a single momentum scale. Our results show that the 
computation of this Green function requires large statistical ensembles with 10K or larger number of gauge configurations. The simulations considered 
herein have a clear Monte Carlo signal for momenta up to $\sim 1$ GeV. The form factors show an hierarchy, with the form factor
associated with the tree level Feynman rule being dominant and essentially constant for 
the range of momenta accessed. The remaining form factors seem to increase as the momentum decreases, suggesting that a possible
$\log$ divergence may  occur. The computed form factors are, at least, in qualitative agreement with the results
obtained with continuum approaches to this vertex, when available.
\end{abstract}

\maketitle

\tableofcontents

\section{Introduction}

The dynamical content of the QCD Green functions is of upmost importance to all hadronic physics, specially its non-perturbative contents. 
Indeed, among other phenomena, the interpretation of the hadron spectra, the comprehension of dynamical chiral symmetry breaking and of
the quark and gluon confinement mechanisms are outside the scope of the perturbative solution of the theory.

The analysis of the QCD Green functions started looking at those functions with smaller number of external legs, namely the propagator or
two point correlation functions. Their tensorial structure is simpler, as it requires smaller tensor basis, and can be guided with the help of the 
Slavnov-Taylor identities. All the two point QCD correlation functions have been studied in detail with non-perturbative methods and
a fairly good understanding of these functions has been achieved
\cite{Cucchieri:2007md,Aguilar:2008xm,Fischer:2008uz,Bogolubsky:2009dc,Oliveira:2012eh,Ayala:2012pb,Duarte:2017wte}. 
A good and coherent picture of the two point QCD correlation functions, obtained with different non-perturbative methods, 
has been achieved but there are issues that are still under debate.
An example is their analytical structure, required in the computation of meson and baryon properties with continuum methods, that
remains to be understood \cite{Alkofer:2003jj,Strauss:2012dg,Falcao:2020vyr,Falcao:2022gxt,Boito:2022rad}.

In QCD the three point one-particle irreducible (1-PI) Green functions have also been object of study. Indeed,
the ghost-gluon vertex
\cite{Ilgenfritz:2006he,Schleifenbaum:2004id,Cucchieri:2008qm,Huber:2012kd}, 
the three gluon vertex
\cite{Binosi:2013rba,Boucaud:2013jwa,Eichmann:2014xya,Athenodorou:2016oyh,Duarte:2016ieu,Rodriguez-Quintero:2017phk,Binosi:2017rwj,Vujinovic:2018nqc,Aguilar:2019jsj,Aguilar:2019uob,Maas:2020zjp,Aguilar:2021lke,Figueroa:2021sjm,Catumba:2021yly,Barrios:2022hzr,Aguilar:2023qqd,Pinto-Gomez:2022brg,Pinto-Gomez:2023zvj,Aguilar:2023mdv}
and the quark-gluon vertex
\cite{Davydychev:2000rt,Bender:2002as,Skullerud:2002ge,Skullerud:2003qu,Bhagwat:2004kj,Kizilersu:2006et,Matevosyan:2007cx,Alkofer:2008tt,He:2009sj,Rojas:2013tza,Aguilar:2014lha,Pelaez:2015tba,Binosi:2016wcx,Bermudez:2017bpx,Oliveira:2018fkj,Oliveira:2018ukh,Sultan:2018tet,Kizilersu:2021jen,Gao:2021wun,Skullerud:2021pel,El-Bennich:2022obe,Marques:2023cmi,Aguilar:2023mam}
have been investigated with some degree of detail and its main features are now understood. 
If for some of the vertices a qualitative picture of their dependence for different kinematics can be claimed,
a complete and accurate description is still to be done.

The next level of QCD one-particle irreducible Green functions to be studied with non-perturbative methods are those with four external legs.
These include the four quark and the four gluon correlation functions. The four quark Green function is required to understand, for example, the meson spectra.
The four gluon correlation function has a similar role as the four quark function but for the glueball spectra. Moreover, it also allows for the definition
of a renormalization group invariant QCD charge, it contributes to the Dyson-Schwinger equation for the gluon propagator and impacts on the gluon propagator itself \cite{Eichmann:2021zuv}. Herein, our focus of attention is the four gluon one-particle irreducible Green function.

The non-perturbative content of the four gluon 1-PI Green function has been studied with the continuum formulation of QCD using the Dyson-Schwinger
equations \cite{Kellermann:2008iw,Binosi:2014kka,Cyrol:2014kca}. A full description of this four leg function requires a tensor
basis with a large number of elements \cite{Gracey:2014ola,Binosi:2014kka,Eichmann:2015nra}, that makes the analysis of this four point 
correlation function difficult. Continuum studies of  this function have to make simplifications to arrive at a manageable computational scenario
and the calculations performed have to consider a basis with a reduced number of elements. 
The continuum based Dyson-Schwinger calculations identify a number of form factors that display very mild dependence with the momentum, 
as seems to be 
the case of the form factor associated with the tree level tensor structure that, nevertheless, seems to be suppressed at low momenta, 
and others that show $\log$ divergences in the IR region. These divergences are associated with the unprotected $\log$'s coming from the massless
ghost loops in the Dyson-Schwinger equations \cite{Binosi:2014kka}. The continuum QCD methods based on the Dyson-Schwinger equation for this
1-PI Green function seem to favour such type of scenario.

The picture that emerges from the various continuum approaches to the four gluon 1-PI Green function, although being consistent at the 
qualitative level,  also show differences that probably come from the different truncations and/or the required modelling 
 to solve the underlying equations. The differences between the various studies provide an extra motivation to a lattice 
calculation of the four gluon vertex.  Indeed, the interplay of the various non-perturbative methods is important to achieve a proper and sufficiently 
accurate description of the QCD one-particle irreducible Green function.

In this work we address the computation of the four gluon one-particle irreducible Green function within pure Yang-Mills lattice QCD simulations 
in the Landau gauge. Preliminary results can be found in \cite{Colaco:2023qin}.
The manuscript is organized as follows. 
In Sec. \ref{Sec1} it is given a description of the four gluon full Green function, accessed in the lattice QCD
simulations, and how it relates to the corresponding 1-PI function. Furthermore, details on the extraction of the 1-PI contribution
in a simulation in the Landau gauge are also discussed there, together with the tensor basis that will be used for the calculation.
The definitions of the form factors to be measured can also be found in this section. 
The lattice setup, the choice of the kinematic configurations to be used in the measurements and some details of the lattice calculation are given in
Sec. \ref{Sec:LatticeSetup}. The lattice form factors, including those associated with the amputated Green functions, are reported and discussed in
Sec. \ref{Sec:FF}. Finally in Sec. \ref{Sec:Final} we summarize and conclude. Some of the details of the calculation, namely a proof of the decoupling
of the three gluon contribution to the four point function for a certain class of kinematical configuration and the lattice version of the continuum
operators, are given in the appendices.

\section{The Four Gluon Vertex \label{Sec1}}

Let us start by discussing the computation of the four gluon 1-PI Green function using the continuum formulation of QCD in Minkowski
spacetime. By doing so we aim to arrive at a better understanding of this complex Green function before plugging into the measurement of the
Green function with lattice QCD simulations.

\begin{figure}[t] 
   \centering
   \includegraphics[width=7in]{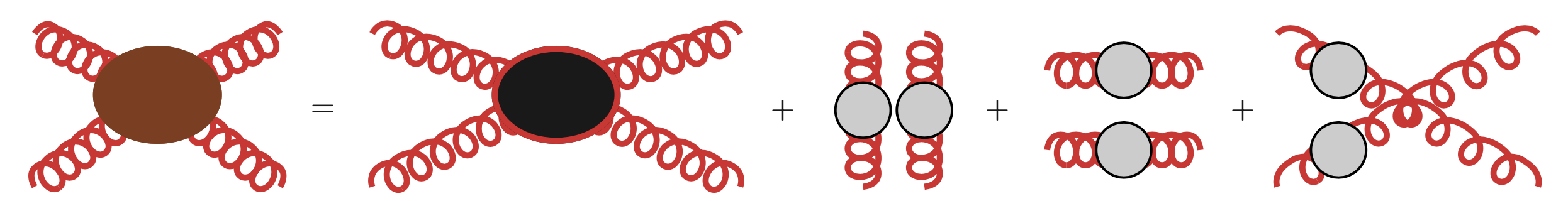} 
\caption{Diagramatic representation of the four gluon full Green function  defined in Eq (\ref{GF4:Complete}). 
The filled vertex in the left is the complete four point Green function, the blob filled in back is the connected Green function and the blobs in grey stand 
for one particle-irreducible diagrams. Color, Lorentz and momentum indices are omitted.}
\label{Fig:completeGF}
\end{figure}

\begin{figure}[t] 
   \centering
   \includegraphics[width=7in]{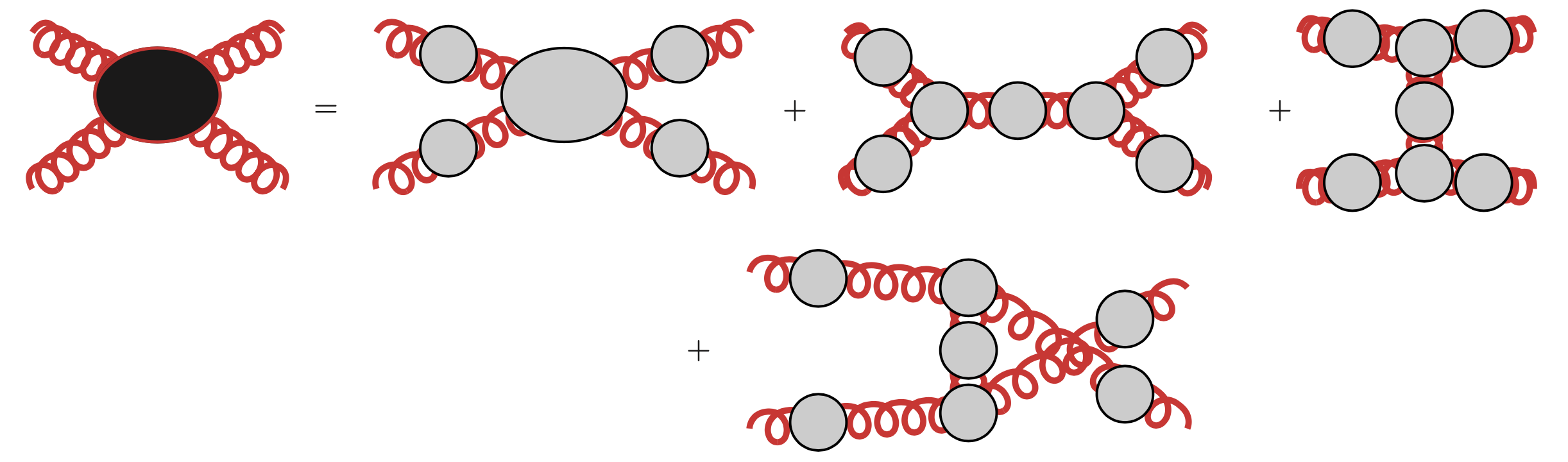} 
\caption{Diagramatic representation of the four gluon connected Green function (full black blob)
appearing in Fig. \ref{Fig:completeGF} in terms of one particle-irreducible functions (full grey blobs).}
\label{Fig:Connected}
\end{figure}

In the numerical Monte Carlo simulations of QCD on a lattice only the full Green functions are accessed. In real space, these Green
functions are defined as the vacuum expectation values of the time order product of the fundamental fields. In QCD, the gluon 
N-point full Green function reads
\begin{equation}
 \mathcal{G}^{a_1 \cdots a_N}_{\mu_1 \cdots \mu_N} (x_1, \, \cdots, \, x_N) = 
  \langle ~ A^{a_1}_{\mu_1} (x_1) ~ A^{a_2}_{\mu_2} (x_2)  ~\cdots ~ A^{a_N}_{\mu_N} (x_N)~\rangle \ ,
  \label{Def:GreenFunctionN}
\end{equation}
where $\langle \cdots \rangle$ represent the vaccuum expectation value, realised as an ensemble average over gauge configurations
sampled with an appropriate action. Any other quantity, like form factors, have to be computed \textit{a posteriori} by handling a set of full 
Green functions $\mathcal{G}^{a_1 \cdots a_N}_{\mu_1 \cdots \mu_N}$. The computation of the 1-PI four gluon Green function is no
exception and the first step to perform is to decompose the 4-point full Green function in terms of  QCD 1-PI Green functions.
For pure Yang-Mills theory the functional generator of the full Green functions $Z[J]$ is
\begin{equation}
Z[J] = e^{ i \, W[J]} = \int \mathcal{D}A \, e^{ i \, S[A] + i (J,A) } \ ,
\end{equation}
where $S[A]$ is the effective action associated with the Yang-Mills theory, that has contributions coming from the gauge and ghost fields,
\begin{equation}
 (J,A)  = \int d^4x ~A^a_\mu(x) J^{a \, \mu}(x) \ ,
\end{equation}
$W[J]$ is the generating functional of the connected Green functions and $J$ are classical external sources. 
It follows from the definition of the full Green function that the four gluon full Green function is given by
\begin{eqnarray}
 \mathcal{G}^{a_1 \cdots a_4}_{\mu_1 \cdots \mu_4} (x_1, \, \cdots, \, x_4) & = &   \left. \frac{\delta^4 Z[J]}{i \, \delta J^{a_1 \, \mu_1} (x_1) ~ \cdots ~ i \, \delta J^{a_4 \, \mu_4}(x_4)} \right|_{J=0} \nonumber \\
 & = & 
      \left. i \, \frac{\delta^4 W[J]}{\delta J^{a_1 \, \mu_1} (x_1) ~ \cdots ~  \delta J^{a_4 \, \mu_4} (x_4)} \right|_{J=0} \nonumber \\
      & & ~ -
     \left. \frac{\delta^2 W[J]}{\delta J^{a_1 \, \mu_1} (x_1) ~  \delta J^{a_2 \, \mu_2} (x_2)} \right|_{J=0} ~ \left. \frac{\delta^2 W[J]}{\delta J^{a_3 \, \mu_3} (x_3) ~  \delta J^{a_4 \, \mu_4} (x_4)} \right|_{J=0} \nonumber \\
  & & \quad\quad - 
     \left. \frac{\delta^2 W[J]}{\delta J^{a_1 \, \mu_1} (x_1) ~  \delta J^{a_3 \, \mu_3} (x_3)} \right|_{J=0} ~ \left. \frac{\delta^2 W[J]}{\delta J^{a_2 \, \mu_2} (x_2) ~  \delta J^{a_4 \, \mu_4} (x_4)} \right|_{J=0} \nonumber \\
  & & \qquad\qquad - 
     \left. \frac{\delta^2 W[J]}{\delta J^{a_1 \, \mu_1} (x_1) ~  \delta J^{a_4 \, \mu_4} (x_4)} \right|_{J=0} ~ \left. \frac{\delta^2 W[J]}{\delta J^{a_2 \, \mu_2} (x_2) ~  \delta J^{a_3 \, \mu_3} (x_3)} \right|_{J=0}  \ .
     \label{GF4:Complete}
\end{eqnarray}
The diagrammatic representation of $\mathcal{G}^{a_1 \cdots a_4}_{\mu_1 \cdots \mu_4} (x_1, \, \cdots, \, x_4)$ in terms of connected Green functions
is given in Fig. \ref{Fig:completeGF}. 
The four gluon full Green function is a sum of a four-gluon connected Green function, represented by the full black blob in Fig. \ref{Fig:completeGF},
and disconnected parts that are functions of the two gluon Green function, i.e. of the gluon propagator, the full blobs in grey.
The four gluon connected Green function itself can be written in terms of 1-PI functions.  In order to workout this decomposition 
let us consider the classical Yang-Mills field given by
\begin{equation}
   A_{c l, \mu}^a (x) = \frac{\delta W[J]}{\delta J^{a \, \mu} (x)} \ 
   \label{Eq:ClassicalField}
\end{equation}
and introduce the generating functional for the one-particle irreducible correlation functions
\begin{equation}
  \Gamma [ A_{cl} ] =  W[J] - (J,A_{cl}) \ .
  \label{GammaDef}
\end{equation}  
The one-particle irreducible Green functions are the functional derivatives of the $\Gamma [ A_{cl} ]$ at vanishing classical fields.
It follows from the above definitions that 
\begin{equation}
   \frac{\delta\, \Gamma}{\delta A_{cl , \mu}^a (x)} = - \, J^{a \, \mu} (x) 
   \label{DeltaGammas}
\end{equation}
and, from this relation combined with Eqs (\ref{Eq:ClassicalField}) and (\ref{GammaDef}), the orthogonality relation
\begin{equation}
    \int d^4 z ~~ \frac{\delta^2  W[J] }{\delta J^a_\mu(x) ~~ \delta J^c_\nu(z)} ~~   \frac{\delta^2 \, \Gamma[A_{cl}]}{ \delta A_{cl}^c \,  ^{ \nu } (z) ~~ \delta A_{cl}^b \,  _{ \zeta } (y) } = - \, \delta^{a b} ~g^{\mu\zeta} ~
    \delta (x- y) 
   \label{DeltaOrtho}
\end{equation}
is derived. By taking functional derivatives of Eq (\ref{DeltaGammas}), using the orthogonality relation (\ref{DeltaOrtho}), 
and after some straightforward algebra, one arrives at the equality
\begin{eqnarray}
& & 
 \left. \frac{\delta^4 W[J] }{\delta J^a_\mu(x) ~ \delta J^b_\nu(y) ~ \delta J^c_\iota(z) ~ \delta J^e_\zeta(t)} \right|_{J =0} ~ = ~
 \int d^4 w_1 ~d^4 w_2 ~ d^4 w_3 ~ \Bigg\{  \nonumber \\
 & & \qquad 
   D^{bb^\prime}_{\nu\nu^\prime}(y - w_1)  ~ D^{ee^\prime}_{\zeta\zeta^\prime}(t-w_2)~
                \frac{ \delta^3 W}{\delta J^a_\mu(x) ~ \delta J^c_\iota (z) ~\delta J^{c^\prime}_{\iota^\prime}( w_3)} ~
                \frac{ \delta^3 \Gamma}{\delta A^{b^\prime}_{cl \, \nu^\prime} (w_1) ~ \delta A^{c^\prime}_{cl \, \iota^\prime} (w_3) ~  \delta A^{e^\prime}_{cl \, \zeta^\prime} (w_2) }  \nonumber \\
 & & \qquad + ~
   D^{cc^\prime}_{\iota\iota^\prime}(z - w_1)  ~ D^{ee^\prime}_{\zeta\zeta^\prime}(t-w_2)~
                \frac{ \delta^3 W}{\delta J^a_\mu(x) ~ \delta J^b_\nu (y) ~\delta J^{b^\prime}_{\nu^\prime}( w_3)} ~
                \frac{ \delta^3 \Gamma}{\delta A^{b^\prime}_{cl \, \nu^\prime} (w_3) ~ \delta A^{c^\prime}_{cl \, \iota^\prime} (w_1) ~  \delta A^{e^\prime}_{cl \, \zeta^\prime} (w_2) }  \nonumber \\
 & & \qquad + ~
   D^{cc^\prime}_{\iota\iota^\prime}(z - w_1)  ~ D^{ee^\prime}_{\nu\nu^\prime}(y-w_2)~
                \frac{ \delta^3 W}{\delta J^a_\mu(x) ~ \delta J^b_\zeta (t) ~\delta J^{b^\prime}_{\zeta^\prime}( w_3)} ~
                \frac{ \delta^3 \Gamma}{\delta A^{b^\prime}_{cl \, \zeta^\prime} (w_3) ~ \delta A^{c^\prime}_{cl \, \iota^\prime} (w_1) ~  \delta A^{e^\prime}_{cl \, \nu^\prime} (w_2) }  ~ \Bigg\} \nonumber \\
 & &
               +   \int d^4 w_1 ~d^4 w_2 ~ d^4 w_3 ~ d^4 w_4 ~
                   D^{aa^\prime}_{\mu\mu^\prime}(x -w_1) ~  D^{bb^\prime}_{\nu\nu^\prime}(y -w_2) ~  D^{cc^\prime}_{\iota\iota^\prime}(z -w_3) ~  D^{ee^\prime}_{\zeta\zeta^\prime}(t -w_4)  \times \nonumber \\
                   & & \qquad\qquad\qquad\qquad\qquad\qquad \times
                    \frac{ \delta^4 \Gamma}{\delta A^{a^\prime}_{cl \, \mu^\prime} (w_1) ~ \delta A^{b^\prime}_{cl \, \nu^\prime} (w_2) ~  \delta A^{c^\prime}_{cl \, \iota^\prime} (w_3) ~  \delta A^{e^\prime}_{cl \, \zeta^\prime} (w_4) }
                    \label{GF4:Connected}
\end{eqnarray}
whose diagrammatic representation is given in Fig. \ref{Fig:Connected}. 
The connected four gluon Green function is a sum of terms that include the four gluon 1-PI Green function
and contributions proportional to the three gluon 1-PI Green functions.

On a lattice simulation it is the full Green function 
$\mathcal{G}^{a_1 \cdots a_4}_{\mu_1 \cdots \mu_4}$, see  Eq. (\ref{GF4:Complete}) and Fig. \ref{Fig:completeGF}, that is measured.
Its decomposition in terms of connected Green functions calls for the four-gluon connected Green function and disconnected terms that are
proportional to products of  gluon propagators. In the lattice formulation of QCD, the Green functions are proportional to the lattice volume
and, therefore, the terms represented in Fig. \ref{Fig:completeGF} are such that the four gluon connected part is proportional to the lattice volume
$V$, while the terms proportional to the propagators are proportional to $V^2$. The lattice volume dependence of the terms that define
$\mathcal{G}^{a_1 \cdots a_4}_{\mu_1 \cdots \mu_4}$ make it almost impossible to have a good Monte Carlo signal for the connected 
four gluon diagram, unless the contribution of the disconnected parts vanish.  
This can be achieved considering only those kinematical configurations such that the momenta $p_i$ and $p_j$ associated, respectively,
with the external legs $i$ and $j$ are such that $p_i + p_j \ne 0$ for all $i$ and $j$. 
From now on, it will be assumed that these conditions on the momenta of the external legs are fulfilled and, therefore,
the contributions due to the disconnected diagrams can be discarded.

The four gluon connected Green function, represented diagrammatically in Fig. \ref{Fig:Connected}, is given by a term that includes the four gluon 1-PI 
Green function and diagrams that include three gluon 1-PI Green functions and gluon propagators. 
Then, to access the four gluon 1-PI Green function from the connected four gluon 
Green function, the three gluon contributions needs to be known.
However, such information on the three gluon contributions is currently not available. One can either introduce some modelling for the three gluon
1-PI Green function, in such a way that its contribution can be removed, or further constrain the kinematical configuration of the 
external legs to suppress the contributions that call for three-gluon one-particle irreducible diagrams. 
For example, as discussed in \cite{Binosi:2014kka}, this is case when the incoming momenta are 
$p_1 = p$, $p_2 = p$, $p_3 = p$ and $p_4 = - 3 \, p$. A first and  preliminary analysis of this kinematical configuration using lattice QCD 
simulations can be found in \cite{Catumba:2021qbh}.

In general, for the Landau gauge, due to the orthogonality of the gluon propagator, for the kinematical configurations where the momenta of the external legs
are all proportional, i.e. for the class of momenta such that
\begin{equation}
p_1 = p , \qquad p_2 = \eta \, p , \qquad  p_3 = \lambda \, p , \qquad  p_4 = - (1 + \eta + \lambda) \, p \ ,
\label{KinematicalConfiguration}
\end{equation}
where $\eta$ and $\lambda$ are real numbers, the terms in the decomposition of the four gluon connected Green function that require the contribution of the three-gluon vertex vanish. 
The proof of this statement can be found in App. \ref{Sec:3gluoes}. 
Herein, in order to avoid any type of extra modelling, we choose to consider a subset of the kinematical configurations that belong to the class
of momenta defined in Eq. (\ref{KinematicalConfiguration}). Although the modelling of the three gluon 1-PI is avoided, our choice based on the
kinematics constraint the type of information that can be accessed in the simulation.

The choice of momenta is important to access the four gluon 1-PI Green function without extra assumptions. However,
the color-Lorentz structure of this Green function is complex. 
For a general kinematical configuration, the number of tensors required to define a tensor basis to fully describe
the four gluon 1-PI is large. 
For example, for the symmetric point, the four gluon vertex requires more than one hundred different tensors 
for its full description \cite{Gracey:2014ola}. General discussions of the tensor basis for this Green function can be found in 
\cite{Binosi:2014kka,Eichmann:2015nra}. Ideally, one would access each of the form factors that multiply each tensor of the complete
basis for the Green function with the help of a suitable projection operator. However, not only  the lattice simulations convolute the 1-PI functions with the
gluon propagators but also the quality of the signal-to-noise ratio in a Monte Carlo simulation favor certain types of projecting operators. 
In principle,  the signal-to-noise ratio can always be improved by generating larger ensembles of gauge configurations, but a practical computation 
has always a limited statistical ensemble.

A complete tensor basis describing the four gluon 1-PI includes several types of operators that are proportional to the metric tensor and/or momenta.
However, given that in a lattice simulation the  measured function convolutes the four gluon 1-PI with gluon propagators, for the class of kinematical 
configurations considered here that is characterized by a single momentum,  see Eq. (\ref{KinematicalConfiguration}), due to the orthogonality of the 
gluon propagator in the Landau gauge, the tensors proportional to momentum do not contribute to the full Green function that is measured in a 
simulation.  This property simplifies considerably the tensor analysis the Green function and the tensors that contribute can only include the metric
tensor for its Lorentz component and $f_{abc}$ or $d_{abc}$, the fully anti-symmetric and fully symmetric structure constants, and/or $\delta^{ab}$ 
in the description of the color part of the Green function. Then, by imposing Bose symmetry it is possible to build the possible tensor structures
that, for the chosen kinematics, contribute to the Green measured on a simulation. A particular tensor that belongs to the class of allowed operators 
is the tensor structure that appears in the perturbative tree level four gluon Feynman rule
\be
 \widetilde{\Gamma}^{(0)} \,  ^{abcd}_{\mu\nu\eta\zeta}  = 
   f_{abr} f_{cdr} \left( g_{\mu\eta} g_{\nu \zeta} - g_{\mu\zeta} g_{\nu \eta} \right) +
       f_{acr} f_{bdr} \left( g_{\mu\nu} g_{\eta\zeta} - g_{\mu\zeta} g_{\nu \eta}   \right)  +
       f_{adr} f_{bcr} \left( g_{\mu\nu} g_{\eta\zeta}  - g_{\mu\eta} g_{\nu \zeta}  \right) \ .
  \label{Eq:four_glue_proj-tree}
\ee
From this tensor another operator that also contributes to the lattice Green function can be build replacing the fully anti-symmetric SU(3) 
structure constants $f_{abc}$ by the fully symmetric $d_{abc}$ and symmetrizing the Lorentz structure part. 
The corresponding tensor reads
\be
 \widetilde{\Gamma}^{(1)} \,  ^{abcd}_{\mu\nu\eta\zeta}  = 
   d_{abr} d_{cdr} \left( g_{\mu\eta} g_{\nu \zeta} + g_{\mu\zeta} g_{\nu \eta} \right) +
       d_{acr} d_{bdr} \left( g_{\mu\zeta} g_{\nu \eta} + g_{\mu\nu} g_{\eta\zeta} \right)  +
       d_{adr} d_{bcr} \left( g_{\mu\nu} g_{\eta\zeta}  + g_{\mu\eta} g_{\nu \zeta}  \right) \ ,
  \label{Eq:four_glue_proj-2}
\ee
Besides the two above tensors, one may also consider the tensor operator
\be
 \widetilde{\Gamma}^{(2)} \,  ^{abcd}_{\mu\nu\eta\zeta}   = 
\Big(  \delta^{ab}  \, \delta^{cd} + \delta^{ac}  \, \delta^{bd} + \delta^{ad}  \, \delta^{bc}  \Big) ~
\Big(  g_{\mu\nu}  \, g_{\eta\zeta} + g_{\mu\eta}  \, g_{\nu\zeta} + g_{\mu\zeta}  \, g_{\nu\eta} 
\Big) 
\label{Eq:four_glue_proj-3}
\ee
that was mentioned in \cite{Binosi:2014kka}. The three above tensors are not orthogonal to each other and their color-Lorentz scalars are
\begin{eqnarray}
\widetilde{\Gamma}^{(0)} \, \cdot \, \widetilde{\Gamma}^{(0)} & = & 108 \, N^2 \, \Big( N^2 - 1 \Big)  = 7776 \ , \\
\widetilde{\Gamma}^{(0)} \, \cdot \, \widetilde{\Gamma}^{(1)} & = & -36 \, \Big(N^4 - 5 \, N^2 + 4 \Big) = -1440\ , \\
\widetilde{\Gamma}^{(0)} \, \cdot \, \widetilde{\Gamma}^{(2)} & = & 0 \\
\widetilde{\Gamma}^{(1)} \, \cdot \, \widetilde{\Gamma}^{(1)} & = & 12 \left(17 \, N^4 \,  - \, 209 \, N^2 \,  - \, \frac{496}{N^2} \, + \, 688\right) =  \frac{4640}{3} \ ,\\
 \widetilde{\Gamma}^{(1)} \, \cdot \, \widetilde{\Gamma}^{(2)} & = & \frac{288 \left (N^4 \, - \, 5 \, N^2 \, + \,  4\right)}{N}= 3840 \ , \\
 \widetilde{\Gamma}^{(2)} \, \cdot \, \widetilde{\Gamma}^{(2)} & = & 216  \Big(  N^4 \, - \, 1 \Big) = 17280 \ ,
\end{eqnarray}
where the numbers in the r.h.s. refer to the particular case where $N = 3$. In general, the above set of tensors do not define a complete tensor basis 
for the Green function that is measured on a simulation and others should be considered. However, in the current work only their contribution will be measured.

\subsection{An orthogonal set of operators}

Although the operators $ \widetilde{\Gamma}^{(0)}$, $ \widetilde{\Gamma}^{(1)}$ and $ \widetilde{\Gamma}^{(2)}$ are not orthogonal
in the color-Lorentz space, by  a linear combination of these operators it is possible to build three orthogonal operators from 
$\widetilde{\Gamma}^{(i)}$.
Indeed, a straightforward calculation shows that the tensor operators
\begin{eqnarray}
t^{(1)}  \,  ^{abcd}_{\mu\nu\eta\zeta}  & = & \widetilde{\Gamma}^{(0)} \,  ^{abcd}_{\mu\nu\eta\zeta} \ ,   \label{TensorOrtho-1} \\
t^{(2)}  \,  ^{abcd}_{\mu\nu\eta\zeta}  & = & 
         \frac{N^2-4}{3 \, N^2} ~ ~\widetilde{\Gamma}^{(0)} \,  ^{abcd}_{\mu\nu\eta\zeta} ~ + ~ \widetilde{\Gamma}^{(1)} \,  ^{abcd}_{\mu\nu\eta\zeta} 
         ~ - ~ \frac{4 \left(N^2-4\right)}{3 \, N \, \left(N^2+1\right)}
             ~~ \widetilde{\Gamma}^{(2)} \,  ^{abcd}_{\mu\nu\eta\zeta}  \nonumber \\
             & = & \frac{5}{27} ~ ~\widetilde{\Gamma}^{(0)} \,  ^{abcd}_{\mu\nu\eta\zeta} ~ + ~ \widetilde{\Gamma}^{(1)} \,  ^{abcd}_{\mu\nu\eta\zeta} 
         ~ - ~ \frac{2}{9}
             ~~ \widetilde{\Gamma}^{(2)} \,  ^{abcd}_{\mu\nu\eta\zeta} 
         \ ,  \label{TensorOrtho-2} \\
t^{(3)}  \,  ^{abcd}_{\mu\nu\eta\zeta}  & = & \widetilde{\Gamma}^{(2)} \,  ^{abcd}_{\mu\nu\eta\zeta}   \label{TensorOrtho-3}
\end{eqnarray}
are orthogonal in the color-Lorentz space and, therefore,
\begin{equation}
t^{(i)} \cdot t^{(j)} ~  =  ~ \mathcal{N}_j ~~\delta^{ij} \ ,
\end{equation}
where the normalization factors $\mathcal{N}_j$ read
\begin{eqnarray}
 \mathcal{N}_1 & = & 108 \, N^2 \, \Big( N^2 - 1 \Big)  = 7776 \ , \\
 \mathcal{N}_2 & = &  \frac{1280}{3} \ ,\\
 \mathcal{N}_3 & = & 216  \Big(  N^4 \, - \, 1 \Big) = 17280 \ ; \\
\end{eqnarray}
once more the numbers in the r.h.s. are for $N = 3$. 
For a general kinematics, the three tensors $t^{(1)}$, $t^{(2)}$ and $t^{(3)}$ do not define a 
complete basis neither for the full Green function nor for the one-particle function. The exception being the configuration of momenta such that
$p_1 = p_2 = p_3 = p$ and $p_4 = - 3 \, p$ where $t^{(1)}$, $t^{(2)}$ and $t^{(3)}$ describe completely the correlation function
\cite{Binosi:2014kka}.

\subsection{Full Green function and lattice form factors \label{Sec:latticeformfactors}}

For the special kinematics under discussion herein, the full Green functions measured in lattice simulations gets only contributions
from the connected four gluon full Green function. In the Landau gauge, the full Green function can be written as
\begin{eqnarray}
\mathcal{G}^{abcd}_{\mu\nu\eta\zeta} (p_1, \, p_2, \, p_3, \, p_4) & = &
\Big( P_\perp (p_1) \Big)_{\mu\mu^\prime} \, \Big( P_\perp (p_2) \Big)_{\nu\nu^\prime} \, 
\Big( P_\perp (p_3) \Big)_{\eta\eta^\prime} \, \Big( P_\perp (p_4) \Big)_{\zeta\zeta^\prime} ~
D(p^2_1) \, D(p^2_2) \, D(p^2_3) \, D(p^2_4) \nonumber \\
& & ~
\Bigg( F(p^2_1, \dots) \,t^{(1)}  \,  ^{abcd}_{\mu^\prime\nu^\prime\eta^\prime\zeta^\prime}  ~ + ~  
          G(p^2_1, \dots) \,t^{(2)} \,  ^{abcd}_{\mu^\prime\nu^\prime\eta^\prime\zeta^\prime}  ~ + ~  
          H(p^2_1, \dots) \,t^{(3)} \,  ^{abcd}_{\mu^\prime\nu^\prime\eta^\prime\zeta^\prime}   ~ + ~ \cdots \Bigg)
          \label{G:FFdef}
\end{eqnarray}
where $F(p^2_1, \dots)$, $G(p^2_1, \dots)$, $H(p^2_1, \dots)$, etc, are Lorentz scalar form factors, $\cdots$ represent the contribution of
the remaining components of the tensor basis, supposedly but not necessarilly orthogonal to the space spanned by $t^{(1)}$ to $t^{(3)}$, and
\begin{equation}
\Big( P_\perp (p) \Big)_{\mu\nu} = g_{\mu\nu} - \frac{p_\mu \, p_\nu}{p^2}
\end{equation} 
is the orthogonal projector in momentum space that appears in the definition of the Landau gauge gluon propagator
\begin{equation}
   D^{ab}_{\mu\nu} (p) = \delta^{ab} ~ \Big( P_\perp (p) \Big)_{\mu\nu} ~  D(p^2) \ .
\end{equation} 
The 1-PI form factors to be measured from the full Green function are
\begin{eqnarray}
F^{(0)} &  = & - \,
 \widetilde{\Gamma}^{(0)} \,  ^{abcd}_{\mu\nu\eta\zeta}   ~~ \mathcal{G}^{abcd}_{\mu^\prime\nu^\prime\eta^\prime\zeta^\prime} (p_1, \, p_2, \, p_3, \, p_4)
 ~~ g^{\mu\mu^\prime} \, g^{\nu\nu^\prime} \, g^{\eta\eta^\prime} \, g^{\zeta\zeta^\prime} \ ,
 \label{FF:F0} \\
 F^{(1)} & = & - \,
 \widetilde{\Gamma}^{(1)} \,  ^{abcd}_{\mu\nu\eta\zeta}   ~~ \mathcal{G}^{abcd}_{\mu^\prime\nu^\prime\eta^\prime\zeta^\prime} (p_1, \, p_2, \, p_3, \, p_4)
 ~~ g^{\mu\mu^\prime} \, g^{\nu\nu^\prime} \, g^{\eta\eta^\prime} \, g^{\zeta\zeta^\prime} \ ,
 \label{FF:F1}
 \ \\
 F^{(2)} & = & - \,
 \widetilde{\Gamma}^{(2)} \,  ^{abcd}_{\mu\nu\eta\zeta}   ~~ \mathcal{G}^{abcd}_{\mu^\prime\nu^\prime\eta^\prime\zeta^\prime} (p_1, \, p_2, \, p_3, \, p_4)
 ~~ g^{\mu\mu^\prime} \, g^{\nu\nu^\prime} \, g^{\eta\eta^\prime} \, g^{\zeta\zeta^\prime}  \ .
 \label{FF:F2}
\end{eqnarray}
In general, the $F^{(i)}$ are linear combinations of  the form factors $F$, $G$, $H$ $\cdots$ that describe the 1-PI four gluon function.
It follows from the definitions (\ref{FF:F0}) -- (\ref{FF:F2}) that their r.h.s. are symmetric under interchange of any pair of momenta 
and Bose symmetry demands that the $F^{(i)}$ can only depend on the momenta through the combinations
\be
F^{(i)} \equiv F^{(i)} \Big( p^2_1 + p^2_2 + p^2_3 + p^2_4 ~  , ~  (p_1p_2) + (p_1p_3) + (p_1p_4) + (p_2p_3) + (p_2p_4) + (p_3p_4) \Big) \ .
\label{Eq:momentum_dependence}
\ee
For the kinematical configurations investigated in the current work, where all momenta are proportional to each other and there is a unique momentum
scale, one can write that $F^{(i)} \equiv F^{(i)}( p^2 )$ to simplify the notation.

\subsection{Lattice form factors $F^{(i)}$ and 1-PI form factors}

Our primer concern in this work is to compute the $F^{(i)}$. These functions are not the form factors that describe the 1-PI four gluon vertex, see
the definition in Eq. (\ref{G:FFdef}), but are given by  linear combinations of $F$, $G$, $H$, $\dots$ Assuming that the functions $F$, $G$ and $H$
give the dominant contributions to the form factors lattice form factors $F^{(i)}$, then 
for the kinematical configurations $(p_1,  ~~ p_2, ~~ p_3, ~~ p_4) = (0, ~~ p, ~~ p, ~~ -2 p)$ and $(p_1,  ~~ p_2, ~~ p_3, ~~ p_4) = (0, ~~ p, ~~ 2 p, ~~ -3 p)$,
 a straightforward calculation gives
\begin{eqnarray}
F^{(0)} (p^2) & = & 7776 \, F(p^2) \ , \label{Eq:DecompF0-1-PI-zero} \\
F^{(1)} (p^2) & = &  - \, 1440 \, F(p^2) ~ + ~ \frac{1280}{3} \,  G(p^2) ~ + ~ 3840 \, H(p^2) \ , \label{Eq:DecompF1-1-PI-zero} \\
F^{(2)} (p^2) & = &  17280 \, H(p^2)  \label{Eq:DecompF2-1-PI-zero}
\end{eqnarray}
while for $(p_1,  ~~ p_2, ~~ p_3, ~~ p_4) = (p, ~~ p, ~~ p, ~~ -3p)$ one has
\begin{eqnarray}
F^{(0)} (p^2) & = & 3888 \, F(p^2) \ ,  \label{Eq:DecompF0-1-PI} \\
F^{(1)} (p^2) & = & - \, 720 \, F(p^2) ~ + ~ \frac{640}{3} \,  G(p^2) ~ + ~ 2400 \, H(p^2) \ ,  \label{Eq:DecompF1-1-PI} \\
F^{(2)} (p^2) & = & 10800 \,  H(p^2) \ . \label{Eq:DecompF2-1-PI}
\end{eqnarray}
If $F$ and $H$ are directly related to $F^{(0)}$ and $F^{(2)}$, the form factor $G(p^2)$ can be accessed by the linear combination
of all the lattice form factors
\begin{equation}
\frac{5}{27} \, F^{(0)} (p^2) + F^{(1)} (p^2) - \frac{2}{9} \, F^{(2)} (p^2) =  \left\{
\begin{array}{l@{\hspace{0.6cm}}l}
\frac{1280}{3} \,  G(p^2) \ , & \mbox{ all kinematics except } (p_1,  ~~ p_2, ~~ p_3, ~~ p_4) = (p, \, p, \, p, \, -3 p) \\
& \\
\frac{640}{3} \,  G(p^2) \ , & \mbox{ for } (p_1,  ~~ p_2, ~~ p_3, ~~ p_4) = (p, \, p, \, p, \, -3 p) \ .
\end{array}
\right.
\end{equation}
We close this section by recalling the reader that for the kinematics $(p_1,  ~~ p_2, ~~ p_3, ~~ p_4) = (p, \, p, \, p, \, -3 p)$ the three tensor structures considered 
define a complete basis \cite{Binosi:2014kka}.

\section{Lattice setup and choice of momenta \label{Sec:LatticeSetup}}

The main goal of this work is to measure the form factors that describe the four gluon one-particle irreducible correlation function using lattice simulations.
The results described below refer to simulations that use the Wilson action for $\beta = 6.0$. For this bare coupling constant the corresponding 
lattice spacing, measured from the string tension, is $a = 0.1016(25)$ fm or, equivalently, $1/a = 1.943 (48)$ GeV. In the conversion from lattice 
to physical units we use the central values just 
quoted\footnote{Note that there is an uncertainty of  $\sim$2.5\% in the scale setting towards physical momentum. 
However, in the following, the uncertainty associated with $1/a$ will be ignored and only the central value will be used. 
See \cite{Boucaud:2017ksi,Duarte:2017wte} for discussions.}.
The computer simulations were performed using the Chroma \cite{Edwards:2004sx} and PFFT \cite{PippigFFT} libraries and run
on the Navigator supercomputer at the University of Coimbra.

The  form factors $F^{(i)}$ defined in Eqs (\ref{FF:F0}) to (\ref{FF:F2}) were computed with several ensembles of gauge configurations.
Our first try was to consider the ensembles used in \cite{Oliveira:2012eh,Duarte:2016iko,Dudal:2018cli} that have 2000 gauge configurations
for the $64^4$ lattice and 1801 gauge configurations for the $80^4$ lattice,
both rotated to the Landau gauge\footnote{
The details on the sampling and the rotation of the gauge configurations towards the Landau gauge can be found in the above cited works.}. 
For these ensembles of configurations the signal-to-noise ratios are quite poor, preventing from achieving any reliable measurement for the $F^{(i)}$. 
To overcome the problem of the signal-to-noise ratio we have generated 9038 gauge configurations, in the Landau gauge,
for a $32^4$ lattice with the same $\beta$ value, together with 4560 gauge configurations for a $48^4$ lattice and $\beta = 6.0$.
Our choice for $\beta$  relies on the study of the gluon propagator \cite{Oliveira:2012eh} that suggests that for this $\beta$ the lattice spacing effects are 
under control with possible mild volume effects in the IR that show up only in the smallest lattice. From now on, we will refer only to the measurements
with these two latter lattices.

Our choice of momenta to measure the form factors  $F^{(i)}$ considers only a subset of the class of momenta defined in 
(\ref{KinematicalConfiguration}). The subset of momenta was chosen with the aim to minimize the effects due to the breaking of rotational 
symmetry and our choice was to use momenta $p_i$ as close to each other as possible.
The momentum configurations considered are
\be
\begin{array}{l@{\hspace{1cm}}l@{\hspace{1cm}}l@{\hspace{1cm}}l@{\hspace{1cm}}l@{\hspace{1cm}}}
  p_1 =  0, & p_2 = p, & p_3 = p, & p_4 = -2 \, p &  ~ \mbox{ referred as } ~ (0, ~~p, ~~~ \, p, ~~ -2p)  \ , \\
  p_1 =  0, &  p_2 = p, & p_3 = 2p, & p_4 = -3 \, p & ~ \mbox{ referred as } ~  (0, ~~p, ~~ 2p, ~~ -3p) \ , \\
  p_1 =  p, & p_2 = p, & p_3 = p, & p_4 = -3 \, p  & ~ \mbox{ referred as } ~ (p, ~~p, ~~~ \, p, ~~ -3p)  \ .
\end{array}  
 \ee
%
In a simulation that uses hypercubic lattices, the breaking of rotational symmetry implies that the measured form factors $F^{(i)}$ are not only
 functions of $p^2$ but also of the H4  invariants $p^{[n]}$ \cite{Becirevic:1999hj,deSoto:2007ht,Vujinovic:2018nqc,Catumba:2021hcx}.
If for the gluon propagator the building of these invariants and the evaluation of the effects due to the breaking of rotational symmetry
is relatively straightforward, a similar analysis of Green functions with higher number of external legs becomes rather involved.
For the three point functions, the effects due to the breaking of rotational symmetry were discussed 
in \cite{Vujinovic:2018nqc} and for an hypercubic lattice, the breaking of rotational symmetry is minimized 
when one of the momenta is proportional to $( 1, \, 1, \, 1, \, 1)$. 
Similar conclusions have been observed for the two point functions 
\cite{Leinweber:1998uu,deSoto:2007ht,Catumba:2021hcx} 
At least for the two point and three point gluon functions it was observed that the replacement of the naive momentum by an improved momentum,
see Eq. (\ref{Eq:ImprovedMom}) in Sec. \ref{Sec:FF}, seem to handle the effects due to the breaking of rotation symmetry.

For the kinematical configurations under consideration, the form factors $F^{(i)}$ are functions of a single momentum scale.
In order to improve the signal-to-noise ratio, given that the form factors are functions of $p^2$,
in all cases we average over equivalent momenta, including negative momenta, before performing the ensemble averages. 
For example, for each gauge configuration the form factors associated with the momenta
\begin{displaymath}
\begin{array}{lrrrrr@{\hspace{0.4cm}}lrrrrr@{\hspace{0.4cm}}lrrrrr@{\hspace{0.4cm}}lrrrrr@{\hspace{0.4cm}}lrrrrr}
 (& 1,  & 1, & 1, & 1 &)     & (& -1, & 1, & 1, & 1 & )    & (& -1, & -1, & 1, & 1 & )     &  (& -1, & -1, & -1, & 1 &)     &  (& -1, & -1, & -1, & -1 & )    \\
   &     &     &     &     &     & (& 1, & -1, & 1, & 1 & )    & ( & -1, & 1, & -1, & 1 & )     &  (& -1, & -1, & 1, & -1 & )     &  &  &  &  & &    \\
   &     &     &     &     &     & (& 1, &  1, & -1, & 1 & )    & ( & 1, & -1, & -1, & 1 & )     &  (& -1, & 1, & -1, & -1 & )    &  &  &  &  & &     \\
   &     &     &     &     &     & (& 1, &  1, &  1, & -1 & )    & ( & -1, & 1, & 1, & -1 & )     &  (& 1, & -1, & -1, & -1 & )    &  &  &  &  & &     \\
   &     &     &     &     &     &  &     &      &      &     &       & ( & 1, & -1, & 1, & -1 & )     &  &  &  &  &  &     &  &  &  &  & &     \\
   &     &     &     &     &     &  &     &     &      &     &      & ( & 1, & 1, & -1, & -1 & )     &  &  &  &  &  &    &  &  &  &  & &     \\
\end{array}
 \end{displaymath}
are averaged before performing the ensemble averages. By doing so we are assuming that the dependence of the form factors on
the H4 invariants $p^{[n]}$ is small or negligible. 

\section{Lattice form factors \label{Sec:FF}}

\begin{figure}[tp] 
   \centering
   \includegraphics[width=3.5in]{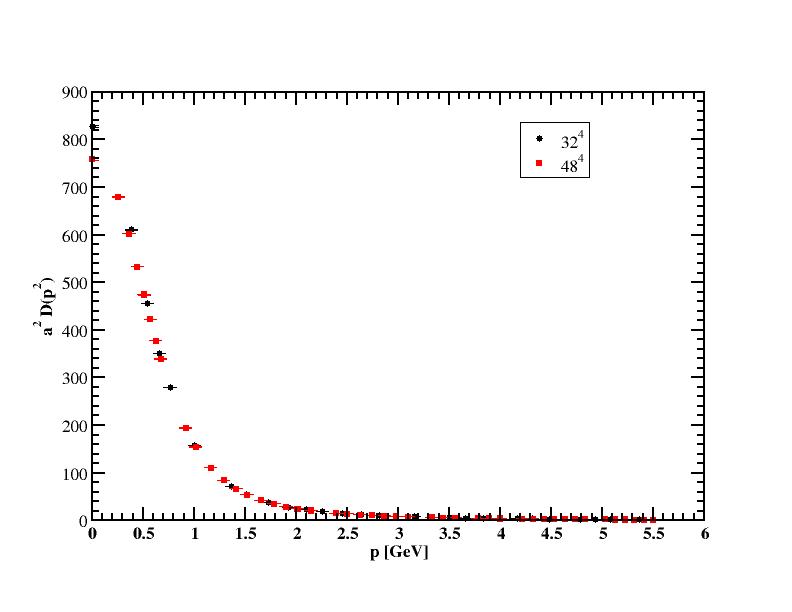} 
   \caption{Bare gluon propagator in the Landau gauge using the data from  the simulations with the $32^4$ and $48^4$ lattices. The data reported 
   is a subset of all the lattice momenta. Indeed,  to handle the effects associated with the breaking of rotational symmetry we follow the recipe of
   \cite{Leinweber:1998uu} and perform the so-called cylindrical and conical cuts in momentum space, while for $p < 0.7$ GeV all data is included.}
   \label{fig:gluon-propagator}
\end{figure}

\begin{figure}[tp] 
   \centering
   \includegraphics[width=2.5in]{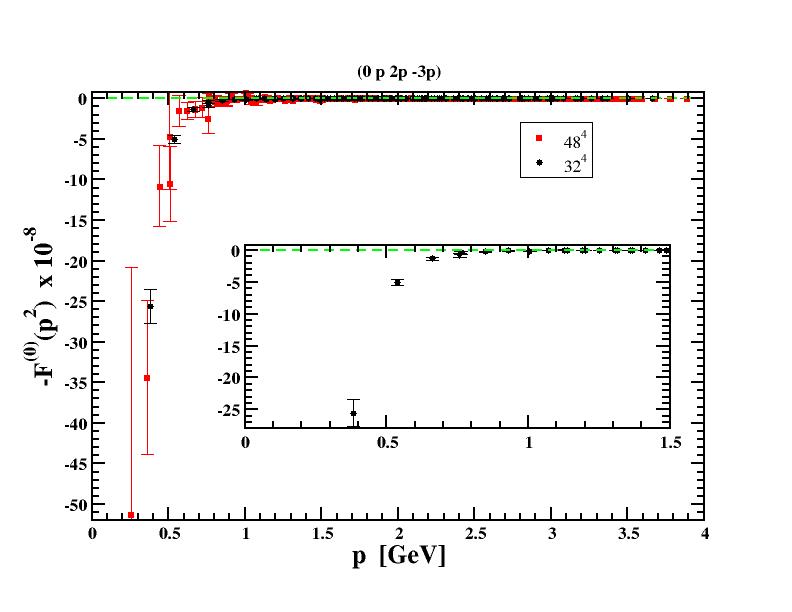} \hspace{-0.8cm}
               \includegraphics[width=2.5in]{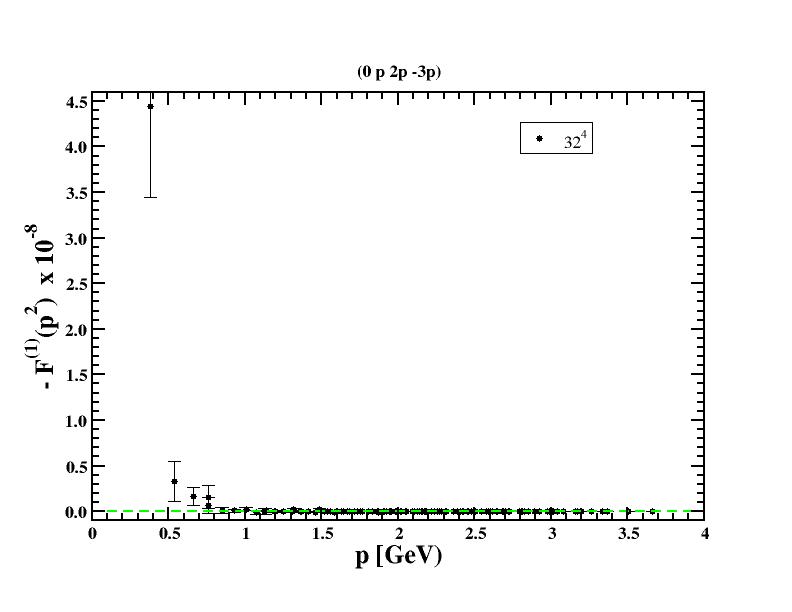}  \hspace{-0.8cm} \includegraphics[width=2.5in]{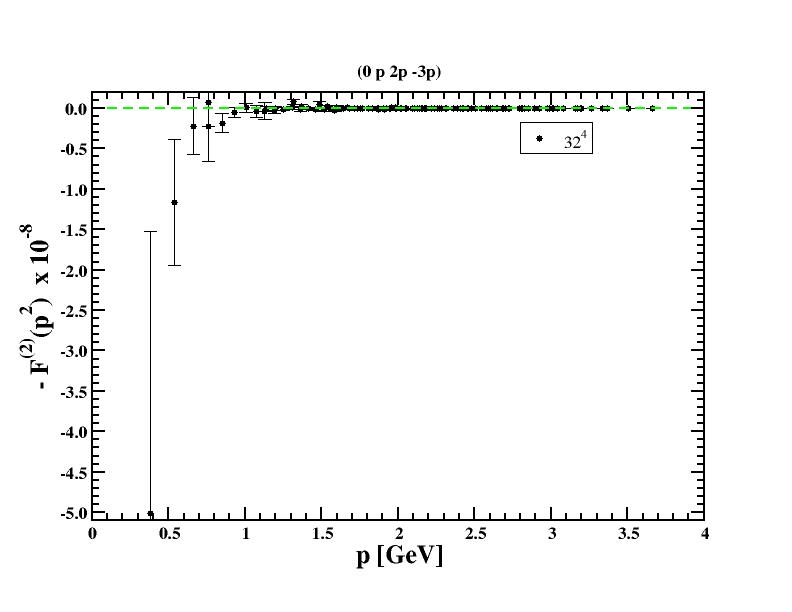}\\
  \vspace{-0.35cm}               
   \includegraphics[width=2.5in]{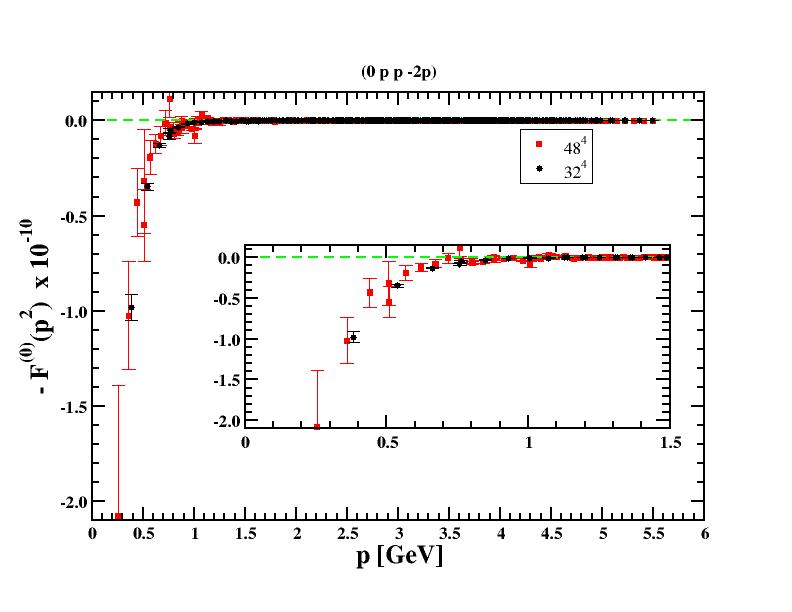} \hspace{-0.8cm}
                \includegraphics[width=2.5in]{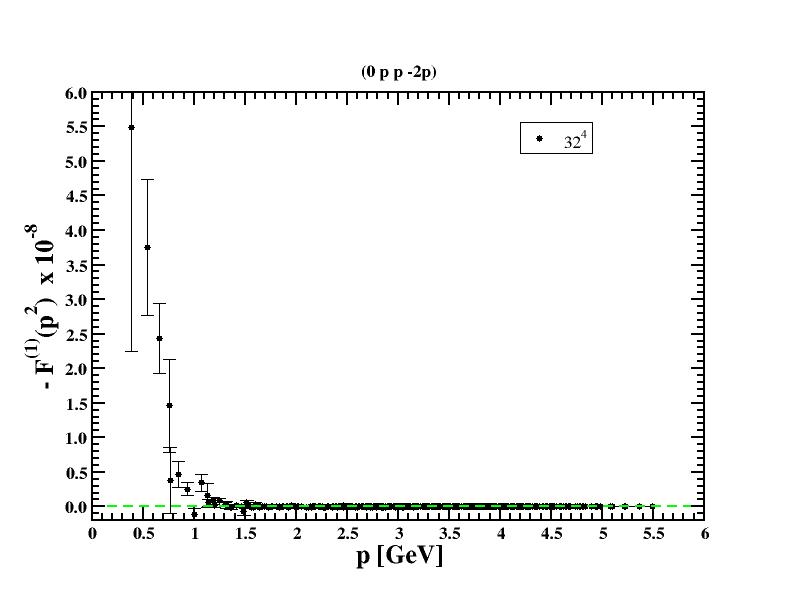} \hspace{-0.8cm} \includegraphics[width=2.5in]{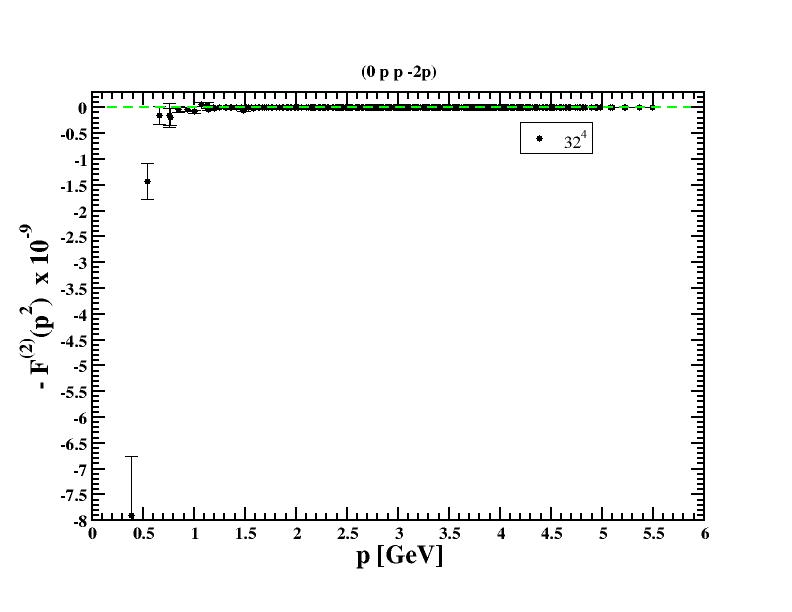}\\
  \vspace{-0.35cm}               
   \includegraphics[width=2.5in]{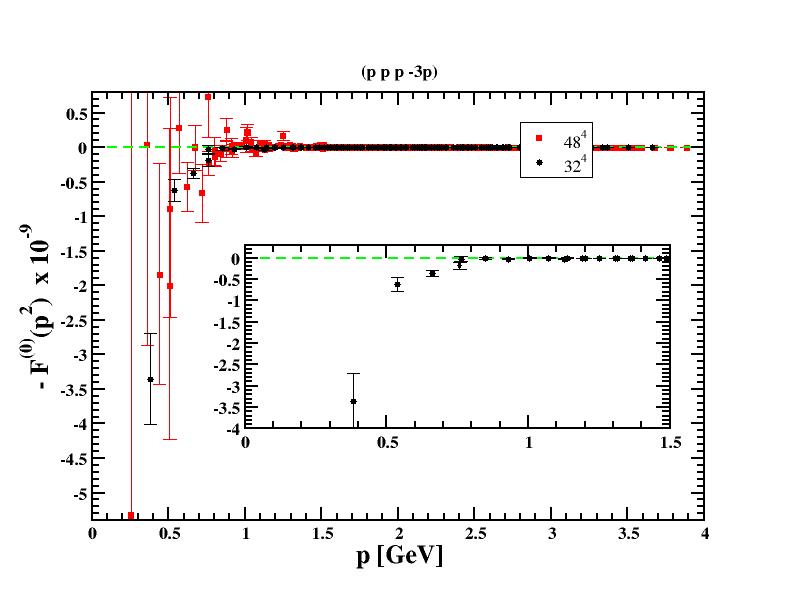} \hspace{-0.8cm}
                \includegraphics[width=2.5in]{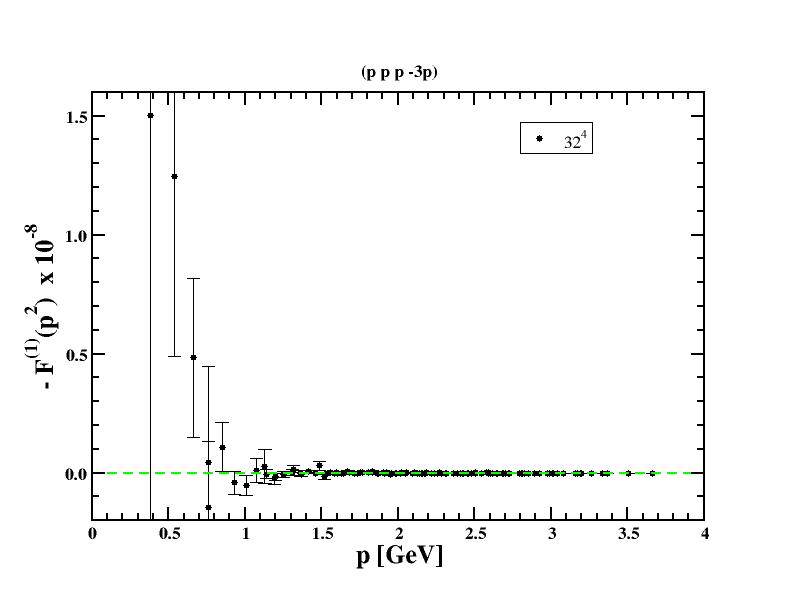} \hspace{-0.8cm} \includegraphics[width=2.5in]{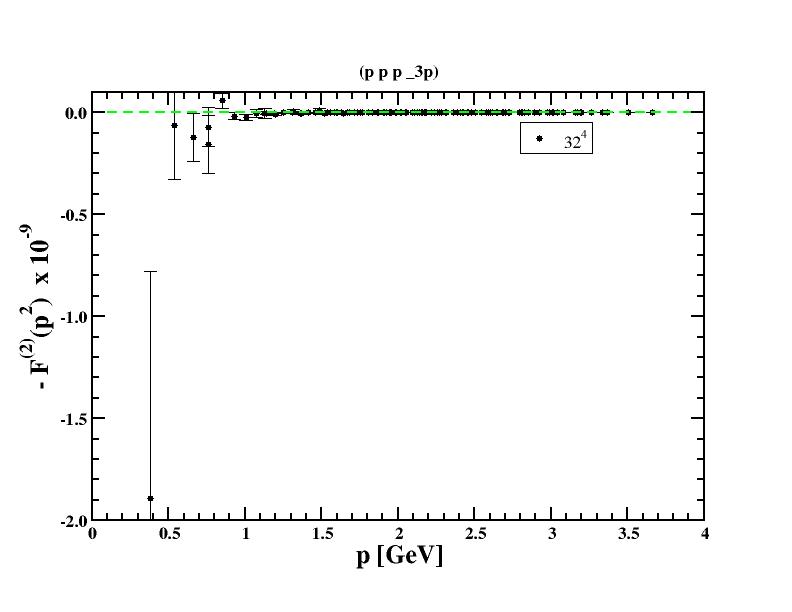}   
   \caption{Dimensionless bare lattice form factors $F^{(i)}$ for the various kinematics. Black dots are the data from the simulations on the $32^4$
   lattice, while the red squares refer to the simulation on $48^4$. Due to the large statistical errors the data from the larger volume is omitted in
   some of the plots. Note the different vertical scales used in each Fig. See main text for details.}
   \label{fig:F0-32}
\end{figure}

The continuum form factors $F^{(i)}$ defined in Eqs (\ref{FF:F0}) to (\ref{FF:F2}) were measured in the simulations using their lattice implementations 
as given in App. \ref{Sec:Momenta}. For the ensembles referred previously, it turns out that only the simulations using the $32^4$ lattice have
a good signal-to-noise ratio on a range of momenta. On the other hand, the results coming from the simulation with the $48^4$ lattice show a 
good signal-to-noise ratio only for $F^{(0)}$ but not for the amputated functions. Anyway, despite the relative large errors associated with this latter lattice, 
the outcome of the simulation confirms the tendency observed in the $32^4$ results and the two sets of data are compatible within one standard deviation.
In all cases, the statistical errors were computed with the bootstrap method with a confidence level of 67.5\%.
The bare lattice data for the form factors  will be described in terms of the improved lattice 
momenta
\begin{equation}
p_\mu = (2/a) \, \sin ( \pi \, n_\mu / L) \qquad\mbox{ with }\qquad n_\mu = -L/2, \, -L/2 +1, \, \cdots, \, 0, \, 1, \, \cdots, \, L/2-1\ ,
\label{Eq:ImprovedMom}
\end{equation}
where $L$ is the length of the lattice. At least for the gluon propagator the use of the improved momenta handles part of the lattice effects.

The bare form factors $F^{(0)}$, $F^{(1)}$ and $F^{(2)}$ associated with the full Green functions, i.e. that  incorporate the contributions of
the gluon propagators, are shown in Fig. \ref{fig:F0-32} for all the kinematics and for all the momenta accessed in the simulation.
The black circles are the data from simulations using the $32^4$ lattice volume, while the red squares is the data measured on the
$48^4$ lattice volume. Note the different vertical scales that are associated with each of the $F^{(i)}$. Their order of magnitude
can be understood looking at the bare gluon propagator function $D(p^2)$, see Fig. \ref{fig:gluon-propagator}, that is included in
the data of Fig. \ref{fig:F0-32}. 

The data reported shows there is a range of momenta where there is a clear sign for all the form factors that goes up to $p \sim 1$ GeV. For higher
momenta the statistical errors prevent the access to the behaviour of the various $F^{(i)}$. 
Reduced errors can be accessed by using larger statistical ensembles of gauge configurations. However, the data 
also shows that the measurement of the four gluon Green functions is possible for gauge ensembles with, at least, about 10k configurations.

 
\begin{figure}[tp] 
   \centering
   \includegraphics[width=2.4in]{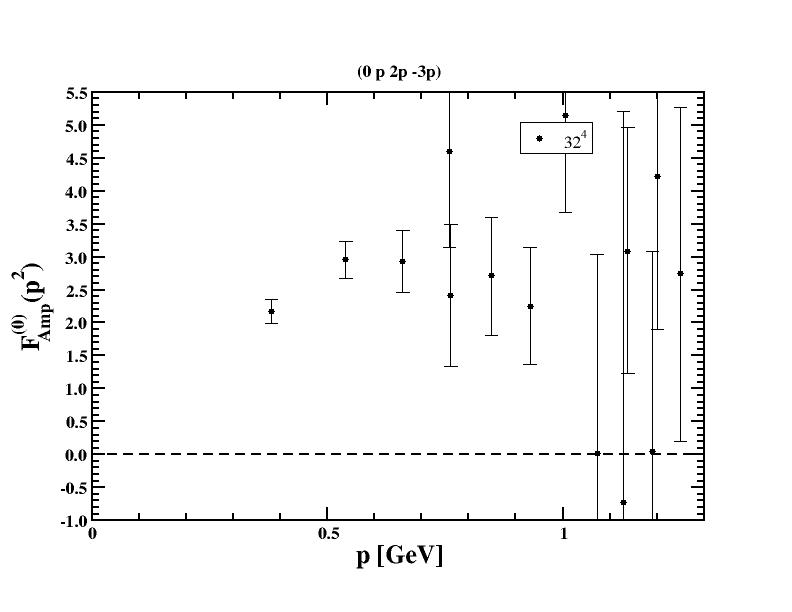} \hspace{-0.8cm}
               \includegraphics[width=2.4in]{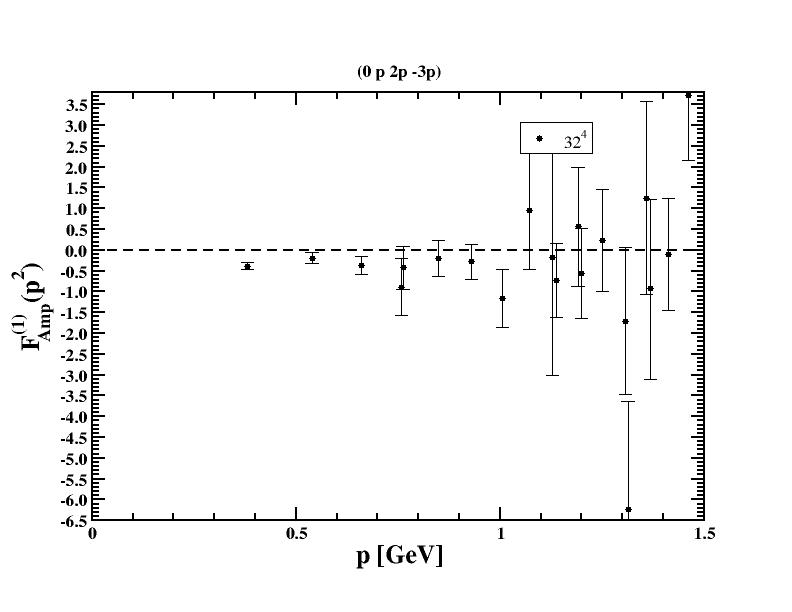}  \hspace{-0.8cm} \includegraphics[width=2.4in]{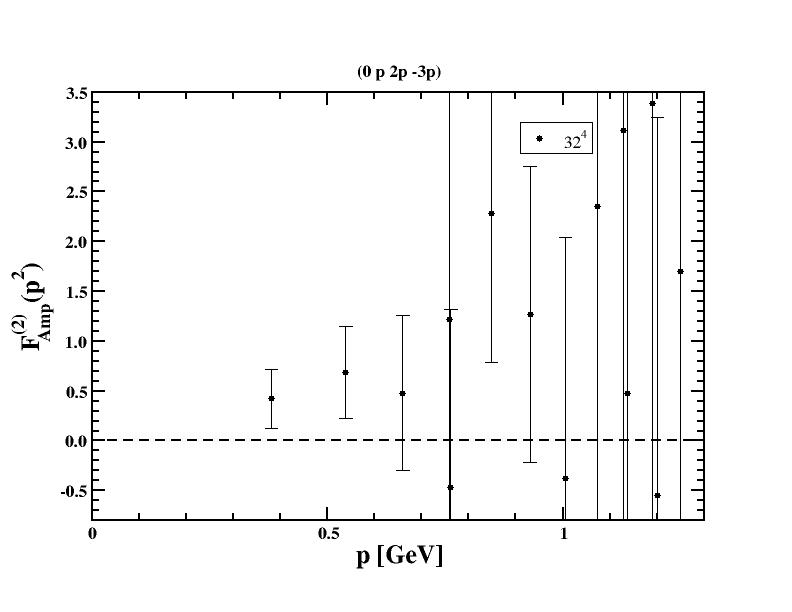}\\
  \vspace{-0.35cm}               
   \includegraphics[width=2.4in]{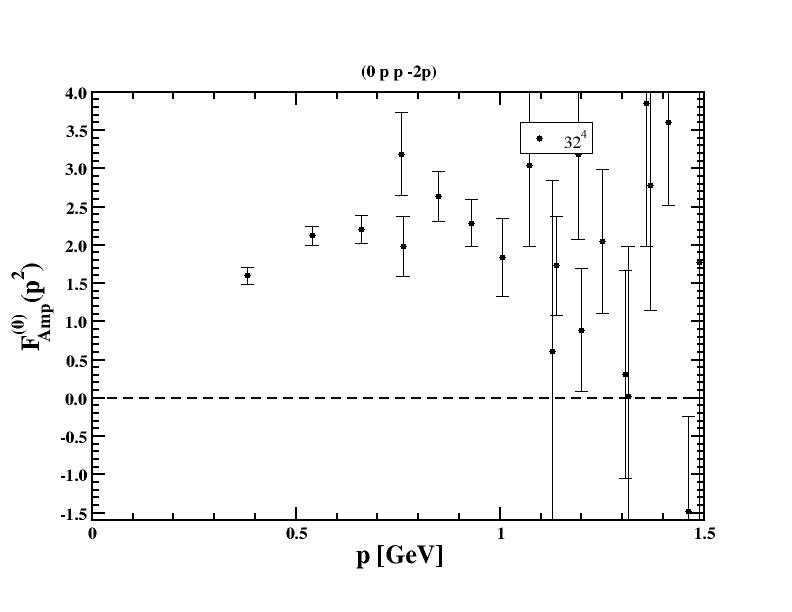} \hspace{-0.8cm}
                \includegraphics[width=2.4in]{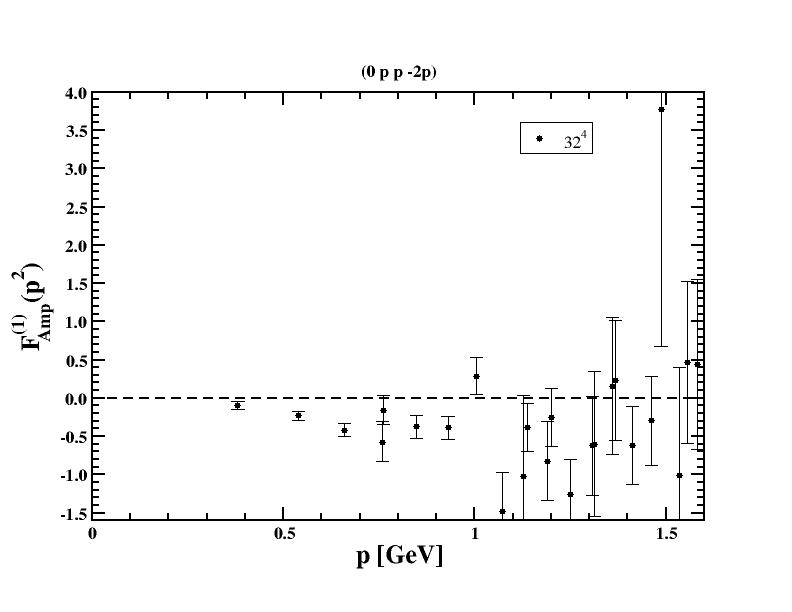} \hspace{-0.8cm} \includegraphics[width=2.4in]{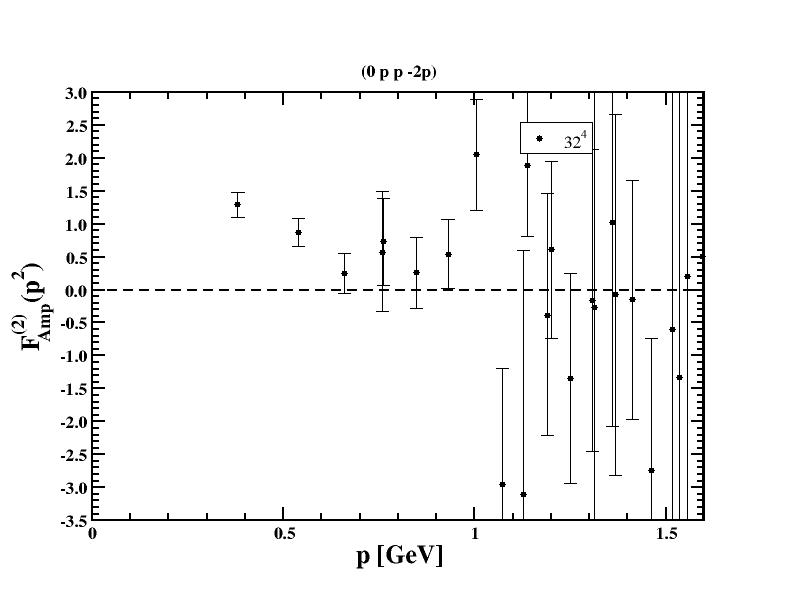}\\
  \vspace{-0.35cm}               
   \includegraphics[width=2.4in]{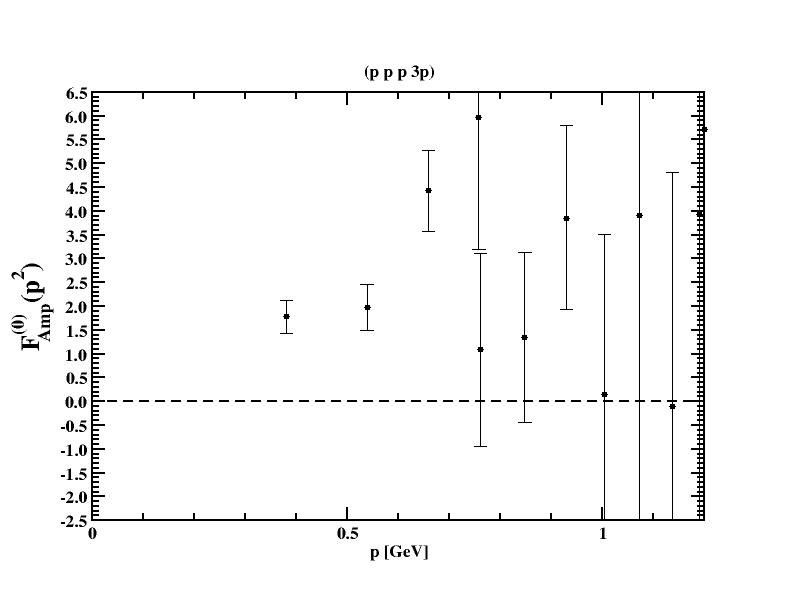} \hspace{-0.8cm}
                \includegraphics[width=2.4in]{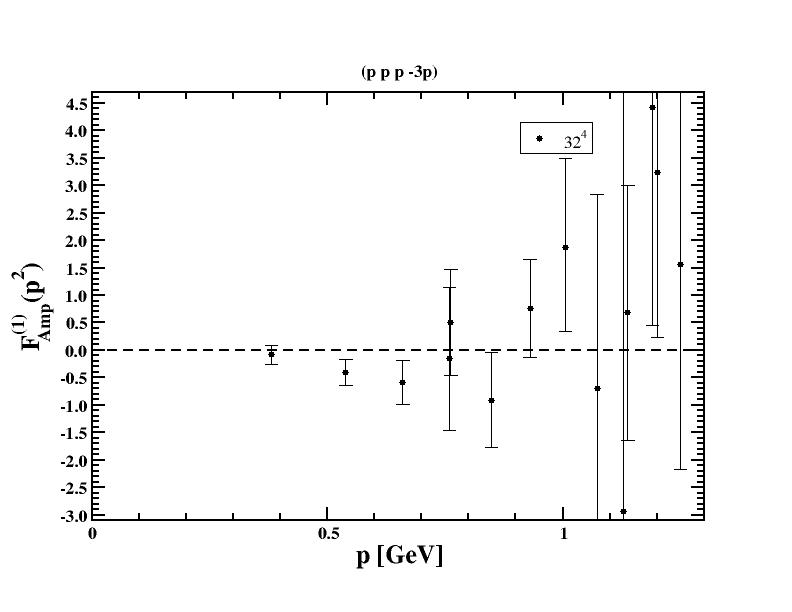} \hspace{-0.8cm} \includegraphics[width=2.4in]{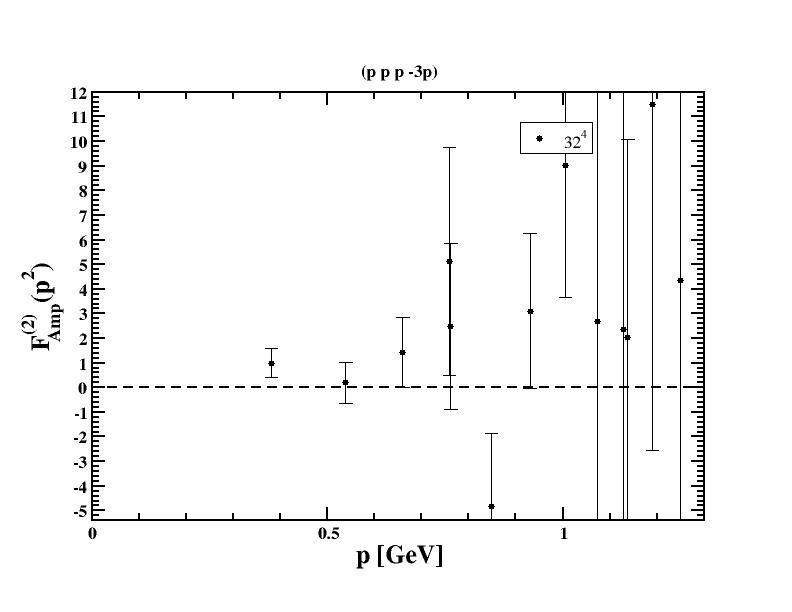}   
   \caption{Dimensionless bare lattice amputated form factors $F^{(i)}$ for the simulation on $32^4$ lattice. The data from the $48^4$ simulation
                  was omitted as it has a two large statistical error to add any useful information. For the lower momenta the data from the simulation with
                  a $48^4$, that have smaller statistical errors, follow the trend observed in the $32^4$ simulation and the two sets of data are
                  compatible within one standard deviation.}
   \label{fig:F0-Amp-32}
\end{figure}

\begin{figure}[tp] 
   \centering
   \includegraphics[width=3in]{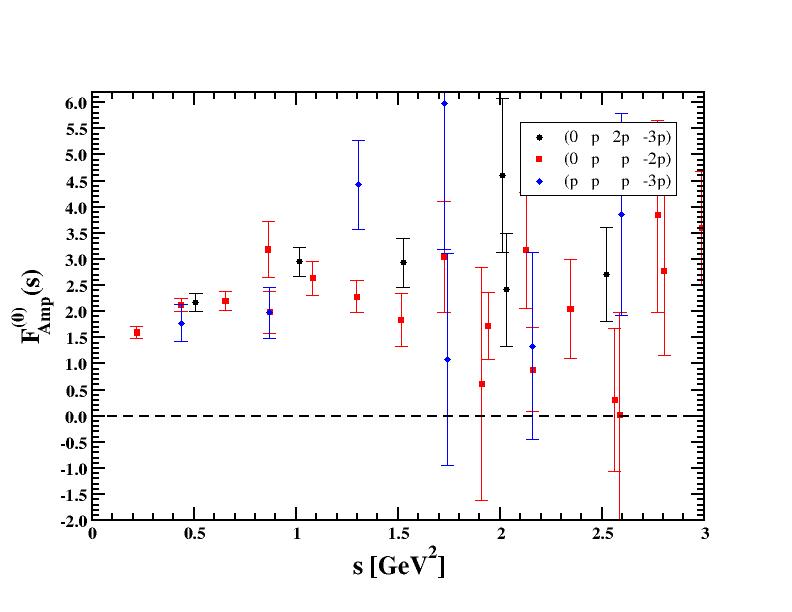} ~ \includegraphics[width=3in]{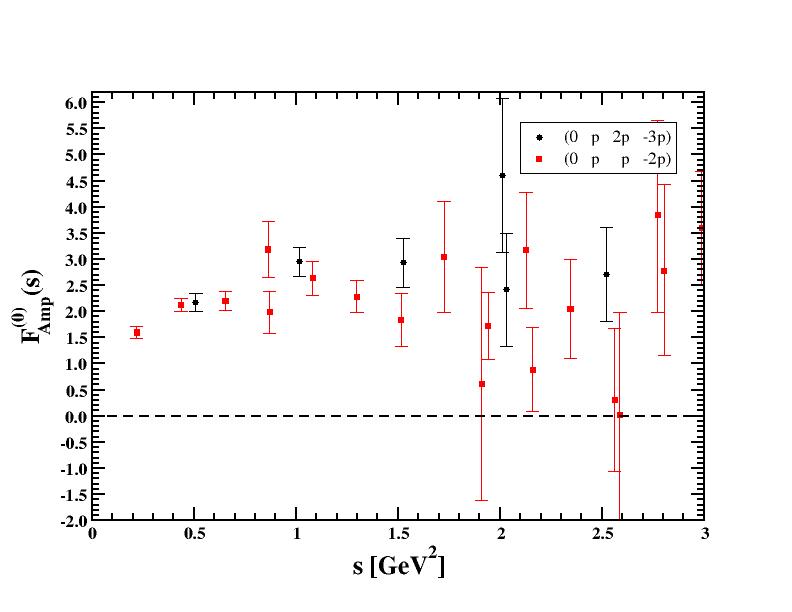} \\
  \vspace{-0.35cm}               
   \includegraphics[width=3in]{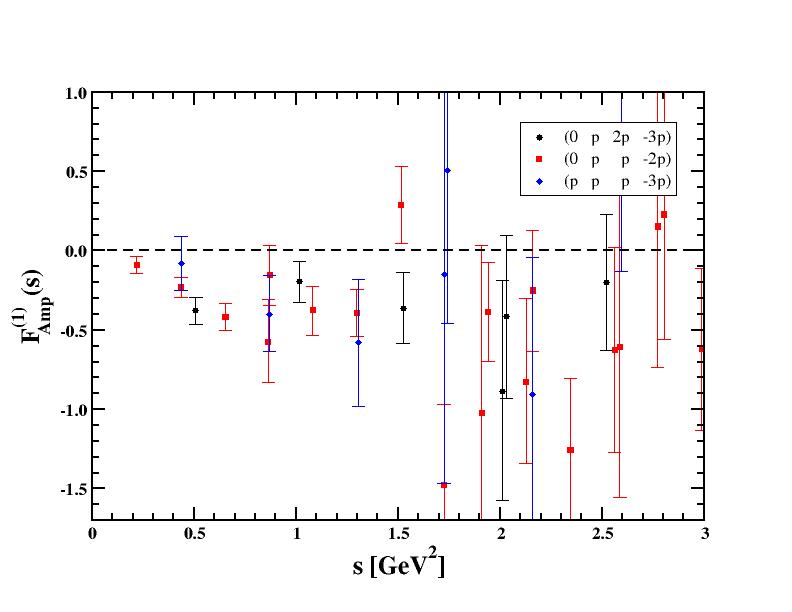} ~ \includegraphics[width=3in]{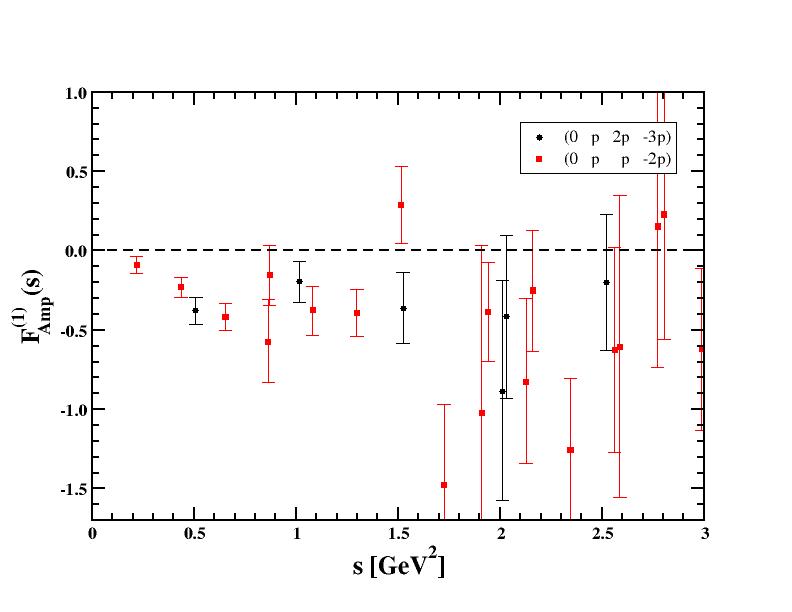} \\
  \vspace{-0.35cm}               
   \includegraphics[width=3in]{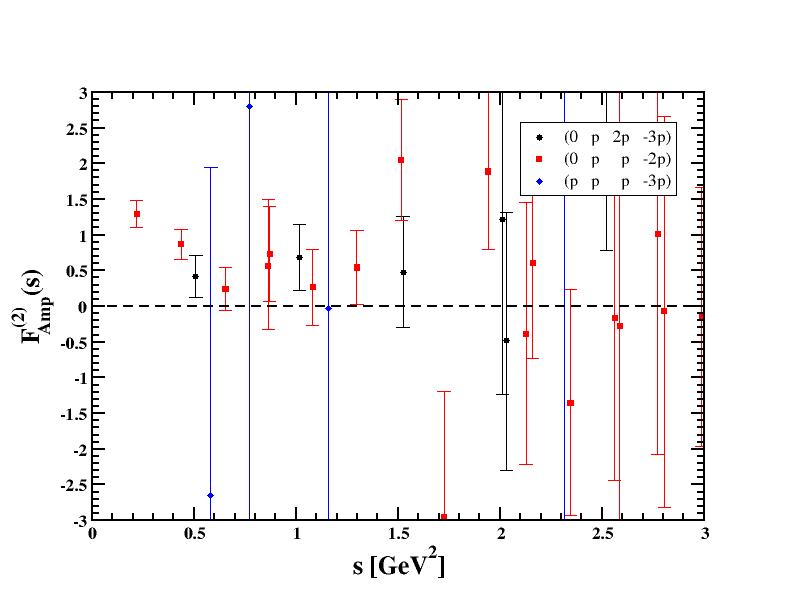} ~\includegraphics[width=3in]{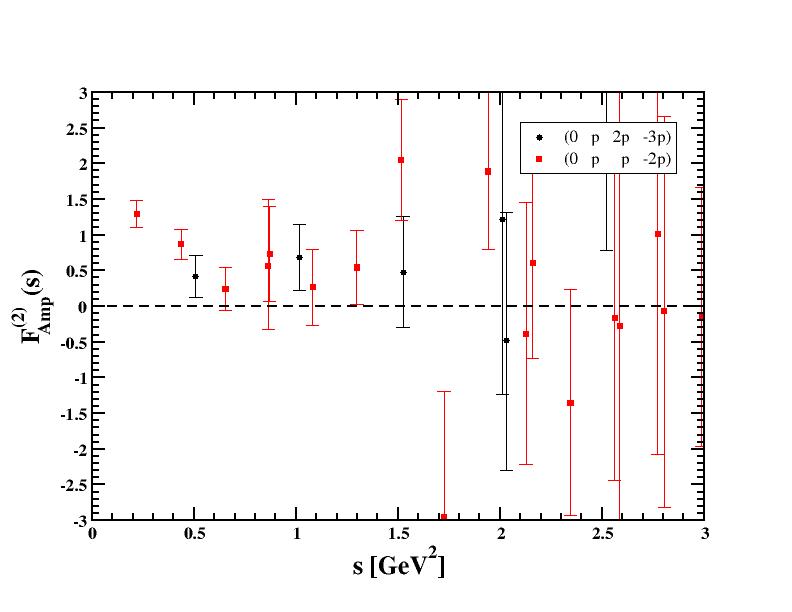} 
   \caption{Dimensionless bare lattice amputated form factors $F^{(i)}$ as a function of $s$ for the simulation on $32^4$ lattice. The plots include
                 the data for all the kinematical configurations. In the right plots the data for kinematical configuration with the large statistical error
                 was omitted for a better reading on the dependence of the form factors with $s$.}
   \label{fig:FF-S}
\end{figure}

Let us now consider the same form factors but associated with the amputated Green functions.
The form factors describing the 1-PI Green function, i.e. the amputated Green function, can be computed from the  $F^{(i)}$ 
dividing the later functions by the gluon propagators that are associated with each of the external legs of the full Green function.
For the size of each gauge ensemble, the statistical errors associated with the two point function are tiny, see the results of
Fig. \ref{fig:gluon-propagator}, and will be ignored when performing the division. Then, the statistical errors for the amputated
Green function form factors come only from the statistical errors on the $F^{(i)}$.

 On a lattice simulation the (bare) computed functions relevant for our purpose are, using now a simplified notation,
 $a^4 \langle A(1) \cdots A(4) \rangle$   for the four point function and $a^2 \langle A(1) A(2) \rangle$ for the propagator.
 From the division of the bare $F^{(i)}$ by the bare the gluon propagator, the would be amputated Green functions are
 \begin{equation}
   \frac{ a^4 \langle A(1) \cdots A(4) \rangle }{\Big( a^2 \langle A(1) A(2) \rangle \Big)^4} =
   \frac{ \langle A(1) \cdots A(4) \rangle }{ a^4 \, \Big( \langle A(1) A(2) \rangle \Big)^4}   \ .
\end{equation}
To arrive at the corresponding dimensionless amputated Green functions, these ratios have to be multiplied by $a^4$.
Then, the bare lattice amputated Green function reads
 \begin{equation}
  a^4 ~  \frac{ a^4 \langle A(1) \cdots A(4) \rangle }{\Big( a^2 \langle A(1) A(2) \rangle \Big)^4}
\end{equation}
and can be seen in Fig. \ref{fig:F0-Amp-32} as a function of the momenta. The data reported shows that there is a clear Monte Carlo
signal
for momenta up to $\sim$ 1 GeV for all the amputated form factors. For larger momenta the statistical errors become quite large and
it is hard to comment on the behaviour of the amputated $F^{(i)}$. 
The dependence on $p$ shows an hierarchy between the various amputated form factors,
with the form factor associated with tree level tensor structure $F^{(0)}$ being the largest of the three $F^{(i)}$. 
In general, for the same form factor, the statistical error associated with kinematical configuration $(p,\, p,\, p, \, -3p)$ is the largest. 
This can be understood looking at the lattice operators given in App. \ref{Sec:Momenta}. Indeed, for this kinematical configuration the 
number of averages performed for each configurations is substantially smaller than for the remaining ones and, therefore, the data associated
with $(p,\, p,\, p, \, -3p)$ should have larger fluctuations, when compared with the other two kinematics investigated here.

In what concerns the dependence of the amputated form factors with momentum, the curves suggest that $F^{(0)}$ is essentially constant for 
$p \gtrsim 0.6$ GeV (although for $p \gtrsim 1$ GeV the large statistical errors prevent any firm conclusion)
and decreases for smaller momenta. The systematics for $F^{(1)}$ and $F^{(2)}$ are harder to understand from 
Fig. \ref{fig:F0-Amp-32}. As discussed below, by taking into account the Bose symmetry and its implication on the amputated form factors
gives a clear understanding on their dependence with momentum scales.

The gluon field is a bosonic field and, therefore, the form factors should depend on momenta as described in
Eq. (\ref{Eq:momentum_dependence}). In this sense, the form factors should be a function not of $p$ but of
\begin{equation}
   s = \left( p^2_1 + p^2_2 + p^2_3 + p^2_4 \right) / 4 
\end{equation}
and of the combination of scalar products of the various momenta mentioned in Eq. (\ref{Eq:momentum_dependence}).  Let us ignore this latter dependence
and redo the plots,  looking at the amputated form factors as a function of $s$ and combining the various kinematics in the same plot. 
The lattice data for the three amputated form factors is reported in Fig. \ref{fig:FF-S}. 
In general, the lattice data for the amputated form factors is compatible with a dependence on $s$. For the level of statistical precision achieved in the
simulation, the description of the lattice does not seem to require any further variable than $s$ itself.
The data of the amputated  $F^{(0)}$ is compatible with a constant for $s \gtrsim 1$ GeV. However, $F^{(0)}$
is slightly suppressed at lower momenta. On the other hand $F^{(1)}$ and $F^{(2)}$ seem to be constant
for $s \gtrsim 1$ GeV but increase when $s$ becomes smaller. 
Moreover, the lattice data for the amputated $F^{(1)}$ suggests that this form factor should change sign at $s \sim 0.3$ GeV. No change of sign
is seen or suggested for $F^{(2)}$. The relative strong increase in the $F^{(1)}$ and $F^{(2)}$ observed when $s$ approaches zero, seem to
suggest that these form factors may have logarithmic divergences in the IR region. However, given the limited access to the low momenta
region, the question of the logarithmic divergences is not solved within the current simulation. To answer this question one should have 
access to the deep infrared region that require simulations using larger lattices or with different lattice spacing.

\section{Summary and Conclusions \label{Sec:Final}}

Herein, the calculation of the four gluon one-particle irreducible Green function, in the Landau gauge, using lattice QCD simulations is addressed.
The determination of the associated form factors from the full Green function is discussed and some of the 1-PI Green function form factors
are computed. Our study shows that lattice simulations to access the four gluon 1-PI Green function require large ensembles of gauge configurations,
meaning 10k or more gauge configurations per ensemble.
Moreover, to resolve the 1-PI contribution to the full four gluon Green function, that is the Green function directly measured in the simulation,
the kinematics of the external legs have to  be carefully chosen. In particular, for the kinematics characterized by a unique momentum, 
it is possible to disentangle the four gluon 1-PI contribution from the disconnected parts and from the diagrams associated with the three gluon 
1-PI in the full Green function.

In order to arrive at such large ensembles of configurations, the simulations reported use either a hypercubic lattice size of $32^4$, 
that has a physical volume of $( \sim 3.3 \mbox{ fm})^4$, or a hypercubic lattice size of $48^4$, whose physical volume is about $( \sim 4.9 \mbox{ fm})^4$.
For the smaller lattice our ensemble has about 10K configurations, for the larger lattice the size of the ensemble is of the order of 4K. 
This prevents the larger lattice volume simulation to achieve a good signal-to-noise ratio and the statistical errors for the 1-PI form factors 
for the $48^4$ lattice are quite large. Nevertheless, they confirm the tendency observed in the simulation with the $32^4$ lattice.
We are currently working on the production of larger statistical ensembles to improve our results, i.e. to achieve a good signal
for larger range of momenta and to explore other kinematical configurations. The results of such simulations will be made public as soon as possible.

The complete description of the four gluon 1-PI irreducible Green function calls for a tensor basis with more than one hundred independent operators. 
However, for the kinematical configurations under consideration, the number of relevant operators that contribute to full Green functions is relatively small
and, in some cases, can be handled exactly. Of the possible tensor operators that define the basis for the 1-PI, we  have considered the three operators
(\ref{Eq:four_glue_proj-tree}), (\ref{Eq:four_glue_proj-2}) and (\ref{Eq:four_glue_proj-3}), measured the associated form factors
$F^{(0)}$,  $F^{(1)}$,  $F^{(2)}$ defined in (\ref{FF:F0}), (\ref{FF:F1}), (\ref{FF:F2}), respectively, and their bare amputated versions that appear in the
1-PI four gluon irreducible Green function.

\begin{figure}[tp] 
   \centering
   \includegraphics[width=6in]{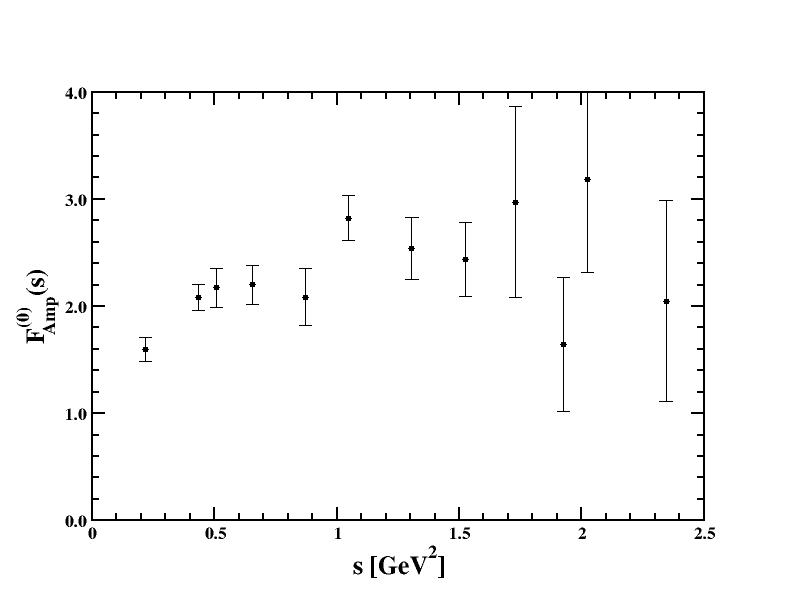} 
   \caption{Dimensionless bare lattice amputated form factors $F^{(0)}$ as a function of $s$ for the simulation on $32^4$ lattice as in Fig. \ref{fig:FF-S}
                replacing the various data points for the same momentum with an weighted average using the statistical error squared as weight.}
   \label{fig:FF-S-Ave}
\end{figure}

The tensor operator used to define the amputated form factor $F^{(0)}$ is the operator that appears in the tree level Feynman rule for the four gluon
vertex. Our simulation gives an amputated form factor that is essentially constant for momentum above $\sim 0.5$ GeV (see Fig. \ref{fig:FF-S}), with
the data suggesting a suppression at small momenta. Despite the large ensemble, for momentum above $\sim 1$ GeV it is difficult to disentangle
its functional form. For momenta in the range $\sim 1 - 1.5$ GeV, despite the size of the statistical errors,
the data seems to be compatible with a constant value; this can be better seen by replacing the various data points for the same momentum
by their weighted averages using the statistical error squared as a weight as reported in Fig. \ref{fig:FF-S-Ave} for $F^{(0)}$. This result is in qualitative
agreement with the outcome of recent Dyson-Schwinger calculations \cite{Binosi:2014kka,Cyrol:2014kca,Colaco:2023qin} but not with
the calculation described in \cite{Kellermann:2008iw}.

The analysis of $F^{(1)}$ shows a negative form factor whose absolute value is about a quarter of the absolute value of $F^{(0)}$, it has the opposite sign of
this latter form factor and seems to approach zero as $s$ is decreased. This behaviour suggests a change of sign for $F^{(1)}$ 
at $s \sim 0.15$ GeV$^2$ and  suggests a divergent behaviour for $F^{(1)}$ in the deep infrared region. Again, the observed 
$F^{(1)}$ is qualitatively in agreement with the Dyson-Schwinger result reported in \cite{Colaco:2023qin}.

The bare amputated $F^{(2)}$ form factor has, apart the possible change of sign, a similar behaviour as observed for $F^{(1)}$, i.e.
$F^{(2)} \sim F^{(0)}/4$, with both form factors having the same sign and with the $F^{(2)}$ data suggesting a possible $\log$ divergence in the deep 
infrared region. The observed lattice data for $F^{(2)}$ is qualitatively in agreement with the recent Dyson-Schwinger calculation
reported in \cite{Colaco:2023qin}. Note that the remaining Dyson-Schwinger studies do not access either $F^{(1)}$ or $F^{(2)}$.

\section*{Acknowledgments}

The authors acknowledge Arlene C. Aguilar, Mauricio N. Ferreira, Joannis Papavassiliou and Leonardo R. Santos for helpful discussions and careful reading
of the manuscript. 
This work was partly supported by the FCT – Funda\c{c}\~ao para a Ci\^encia e a Tecnologia, I.P., under Projects Nos. 
UIDB/04564/2020 (\url{https://doi.org/10.54499/UIDB/04564/2020}), 
UIDP/04564/2020 (\url{https://doi.org/10.54499/UIDP/04564/2020}).
and CERN/FIS-PAR/0023/2021. P. J. S. acknowledges financial support from FCT contract CEECIND/00488/2017 
(\url{https://doi.org/10.54499/CEECIND/00488/2017/CP1460/CT0030}).
The authors acknowledge the Laboratory for Advanced Computing at the University of Coimbra (\url{http://www.uc.pt/lca}) for providing 
access to the HPC resources that have contributed to the research within this paper. Access to Navigator was partly supported by the FCT Advanced Computing 
Project 2021.09759.CPCA.

\appendix

\section{Three gluon contributions for proportional momenta \label{Sec:3gluoes}}

The three-gluon one-particle diagram can be analysed using the Ball-Chiu tensor basis introduced in \cite{Ball:1980ax}; see also \cite{Ahmadiniaz:2012xp}. 
For the momenta configuration of type $p_1 = p$, $p_2 = \eta \, p$ and $p_3 = - ( 1 + \eta) \, p$, only the so called longitudinal component contributes. Following
the notation of Ball-Chiu, the one-particle longitudinal component reads
\bea
 & &
   (1 - \eta) \,  A(p^2_1, \, p^2_2; \, p^2_3) ~ g_{\mu_1 \mu_2} ~ p_{\mu_3}  +
        (1 + 2 \, \eta) \,  A(p^2_2, \, p^2_3; \, p^2_1) ~ g_{\mu_2 \mu_3} ~ p_{\mu_1}  -
        (2 +  \, \eta) \,  A(p^2_3, \, p^2_1; \, p^2_2) ~ g_{\mu_3 \mu_1} ~ p_{\mu_2}     \nonumber \\
 & &  \quad      
  +  ~(1 + \eta) \,  B(p^2_1, \, p^2_2; \, p^2_3) ~ g_{\mu_1 \mu_2} ~ p_{\mu_3}  -
        (1 + 2 \, \eta) \,  B(p^2_2, \, p^2_3; \, p^2_1) ~ g_{\mu_2 \mu_3} ~ p_{\mu_1}  -
         \eta \,  B(p^2_3, \, p^2_1; \, p^2_2) ~ g_{\mu_3 \mu_1} ~ p_{\mu_2}     \nonumber \\
 & &   \quad      
  -  ~ \eta (1 - \eta) \,  C(p^2_1, \, p^2_2; \, p^2_3) ~  p^2 ~P^{\perp}_{\mu_1 \mu_2} (p) ~ p_{\mu_3}  +
        \eta (1 +  \, \eta) \,  C(p^2_2, \, p^2_3; \, p^2_1) ~ p^2 ~ P^{\perp}_{\mu_2 \mu_3} (p) ~  p_{\mu_1}  \nonumber \\
        & & \quad
        - (1 +\eta ) ( 2 + \eta ) \,  C(p^2_3, \, p^2_1; \, p^2_2) ~p^2 ~ P^{\perp}_{\mu_1 \mu_3} (p) ~ p_{\mu_2}  
~   - ~  2 \, \eta \, ( 1 + \eta ) S (p^2_1, \, p^2_2, \, p^2_3) ~ p_{\mu_1} \, p_{\mu_2} \, p_{\mu_3} \, 
\eea
where $A$, $B$, $C$, $S$ are the Ball-Chiu form factors and
\be
 P^{\perp}_{\mu \nu} (p) = g_{\mu\nu} - \frac{p_\mu p_\nu}{p^2} \qquad\mbox{ with }\qquad p^\mu P^{\perp}_{\mu \nu} (p) = 0
\ee
is the orthogonal projector for vector fields. The external legs are proportional to $p$ at least for one of the Lorentz indices,
given that in the contribution to the four-gluon connected Green function all the external legs are contracted with a gluon propagator that, in the Landau gauge,
is proportional to $P^{\perp}_{\mu \nu} (p)$, given that $P^{\perp}_{\mu \nu} (\kappa p) = P^{\perp}_{\mu \nu} (p)$, where $\kappa$ is a constant, 
then it follows that for this class of kinematical configuration the terms in Fig. \ref{Fig:Connected} that contain the three-gluon irreducible diagram vanish.
Note that this result is valid even if one of the momenta vanishes.

\section{Implementation of color-Lorentz projectors to measure $F^{(i)}$ \label{Sec:Momenta}}

The tensor operators that define the lattice form factors $F^{(i)}$, see Eqs (\ref{FF:F0}) to (\ref{FF:F2}), require the SU(3) 
totally antisymmetric group structure constants $f_{abc}$, the totally symmetric structure constants $d_{abc}$ or color diagonal $\delta^{ab}$ 
operators.  The first two types of operators can be computed from color traces of the generators given in the fundamental representation $t^a$ as
\begin{equation}
   \mbox{Tr} \left( t^a \, t^b \, t^c \right) = \frac{1}{4} \big( d_{abc} + i \, f_{abc} \big)
\end{equation}   
or, alternatively, 
\begin{equation}
   - \, 2 \, i \, \mbox{Tr} \left( \left[  t^a \, , \,  t^b  \right] \, t^c \right) =  f_{abc} \qquad\mbox{ and }\qquad
    2 \,  \, \mbox{Tr} \left( \left\{  t^a \, , \,  t^b  \right\} \, t^c \right) =  d_{abc}  \ .
\end{equation}
The Euclidean version of the tensors in Eqs (\ref{Eq:four_glue_proj-tree}) to (\ref{Eq:four_glue_proj-3}) is
\begin{eqnarray}
 \widetilde{\Gamma}^{(0)} \,  ^{abcd}_{\mu\nu\eta\zeta}  & = &
   f_{abr} f_{cdr} \Big( \delta_{\mu\eta} \, \delta_{\nu \zeta}  -  \delta_{\mu\zeta} \, \delta_{\nu \eta} \Big) +
       f_{acr} f_{bdr} \Big( \delta_{\mu\nu} \, \delta_{\eta\zeta} - \delta_{\mu\zeta} \, \delta_{\nu \eta}   \Big)  +
       f_{adr} f_{bcr} \Big( \delta_{\mu\nu} \, \delta_{\eta\zeta}  - \delta_{\mu\eta} \, \delta_{\nu \zeta}  \Big) \ , \\
 \widetilde{\Gamma}^{(1)} \,  ^{abcd}_{\mu\nu\eta\zeta}   & =  &
   d_{abr} d_{cdr} \Big( \delta_{\mu\eta} \, \delta_{\nu \zeta} + \delta_{\mu\zeta} \, \delta_{\nu \eta} \Big) +
       d_{acr} d_{bdr} \Big( \delta_{\mu\zeta} \, \delta_{\nu \eta} + \delta_{\mu\nu} \, \delta_{\eta\zeta} \Big)  +
       d_{adr} d_{bcr} \Big( \delta_{\mu\nu} \, \delta_{\eta\zeta}  + \delta_{\mu\eta} \, \delta_{\nu \zeta}  \Big)  \ , \\
 \widetilde{\Gamma}^{(2)} \,  ^{abcd}_{\mu\nu\eta\zeta}   & = &
\Big(  \delta^{ab}  \, \delta^{cd} + \delta^{ac}  \, \delta^{bd} + \delta^{ad}  \, \delta^{bc}  \Big) ~
\Big(  \delta_{\mu\nu}  \, \delta_{\eta\zeta} + \delta_{\mu\eta}  \, \delta_{\nu\zeta} + \delta_{\mu\zeta}  \, \delta_{\nu\eta} 
\Big) 
\end{eqnarray}
and 
\begin{equation}
F^{(i)}  ~ = ~ \quad
    \widetilde{\Gamma}^{(i)} \,  ^{abcd}_{\mu\nu\eta\zeta}  \quad
    \langle A^a_\mu (p_1) ~ A^b_\nu (p_2) A^c_\eta (p_3) A^d_\zeta (p_4)  \rangle \ .
\end{equation}
Then, plugging in these definitions it turns out that for the (K. Conf. 1) kinematics
\begin{center}
\begin{displaymath}
\boxed{ p_1 = 0, \, p_2 = p_3 = p,  ~ p_4 = - \, 2 \, p }
\end{displaymath}
\end{center}
\begin{eqnarray}
F^{(0)} (p^2) & =  & - \, 8  ~ \sum_r  ~ \Bigg\{
                        \mbox{Tr} \Big( \left[ A_\mu (0) \, , \,  A_\nu (p) \right] \, t^r \Big) \, \mbox{Tr} \Big( \left[ A_\mu (p) \, , \,  A_\nu (-2p) \right] \, t^r \Big) 
                        \nonumber \\
                         & & \hspace{2cm}
                         - \, \mbox{Tr} \Big( \left[ A_\mu (0) \, , \,  A_\nu (p) \right] \, t^r \Big) \, \mbox{Tr} \Big( \left[ A_\nu (p) \, , \,  A_\mu (-2p) \right] \, t^r \Big) 
                        \nonumber \\
                         & & \hspace{3cm} 
                         +\, \mbox{Tr} \Big( \left[ A_\mu (0) \, , \,  A_\nu (- 2 p) \right] \, t^r \Big) \, \mbox{Tr} \Big( \left[ A_\mu (p) \, , \,  A_\nu (p) \right] \, t^r \Big) 
                         \Bigg\} \ , 
 \label{FF:F0:0pp2p}  \\
F^{(1)} (p^2) & = & 8 ~ 
       \sum_r  ~ \Bigg\{ 
                        \mbox{Tr} \Big( \left\{ A_\mu (0) \, , \,  A_\nu (p) \right\} \, t^r \Big) \, \mbox{Tr} \Big( \left\{ A_\mu (p) \, , \,  A_\nu (-2p) \right\} \, t^r \Big) 
                        \nonumber \\
                         & & \hspace{3cm}
                         + \, \mbox{Tr} \Big( \left\{ A_\mu (0) \, , \,  A_\nu (p) \right\} \, t^r \Big) \, \mbox{Tr} \Big( \left\{ A_\nu (p) \, , \,  A_\mu (-2p) \right\} \, t^r \Big) 
                        \nonumber \\
                         & & \hspace{4cm} 
                         +\, \mbox{Tr} \Big( \left\{ A_\mu (0) \, , \,  A_\nu (- 2 p) \right\} \, t^r \Big) \, \mbox{Tr} \Big( \left\{ A_\mu (p) \, , \,  A_\nu (p) \right\} \, t^r \Big) 
                         \Bigg\} \ ,  \\
F^{(2)} (p^2) & = & 4 ~ 
       \Bigg\{   2 \, \, \mbox{Tr} \Big( A_\mu (0) \, \,  A_\mu (p) \Big) \, \mbox{Tr} \Big( A_\nu (p) \, \,  A_\nu (-2p)  \, \Big) \nonumber
       \\
       & & \hspace{2cm}
                 +  2\,\,     \mbox{Tr} \Big( A_\mu (0) \, \,  A_\nu (p) \Big) \, \mbox{Tr} \Big( A_\mu (p) \, \,  A_\nu (-2p)  \, \Big)
                        \nonumber \\
                         & & \hspace{3cm}
                         + \, 2 \,\,  \mbox{Tr} \Big(  A_\mu (0) \,  \,  A_\nu (p) \Big) \, \mbox{Tr} \Big( A_\nu (p) \,  \,  A_\mu (-2p) \Big) 
                         \nonumber \\
                         & & \hspace{4cm}
                         + \, 2 \,\,  \mbox{Tr} \Big(  A_\mu (0) \,  \,  A_\nu (-2p) \Big) \, \mbox{Tr} \Big( A_\mu (p) \,  \,  A_\nu (p) \Big)             
                         \nonumber \\
                         & & \hspace{5cm}
                         + \,  \mbox{Tr} \Big(  A_\mu (0) \,  \,  A_\mu (-2p) \Big) \, \mbox{Tr} \Big( A_\nu (p) \,  \,  A_\nu (p) \Big)             
                         \Bigg\} \ ,
 \label{FF:F2:0pp2p}
\end{eqnarray}
for the (K. Conf. 2) kinematics 
\begin{center}
\begin{displaymath}
\boxed{ p_1 =  0,~  p_2 = p, ~ p_3 = 2 \,p,  ~ p_4 = - \, 3 \, p }
\end{displaymath}
\end{center}
\begin{eqnarray}
F^{(0)} (p^2) & =  &  \, - \, 4 \,  ~ \sum_r  ~ \Bigg\{
                        \mbox{Tr} \Big( \left[ A_\mu (0) \, , \,  A_\nu (p) \right] \, t^r \Big) \, \mbox{Tr} \Big( \left[ A_\mu (2p) \, , \,  A_\nu (-3p) \right] \, t^r \Big)  
                        \nonumber \\
                        & & \hspace{2.5cm} - ~
                        \mbox{Tr} \Big( \left[ A_\mu (0) \, , \,  A_\nu (p) \right] \, t^r \Big) \, \mbox{Tr} \Big( \left[ A_\nu (2p) \, , \,  A_\mu (-3p) \right] \, t^r \Big)  
                        \nonumber \\
                        & & \hspace{3cm} + ~
                        \mbox{Tr} \Big( \left[ A_\mu (0) \, , \,  A_\nu (2p) \right] \, t^r \Big) \, \mbox{Tr} \Big( \left[ A_\mu (p) \, , \,  A_\nu (-3p) \right] \, t^r \Big)  
                        \nonumber \\
                        & & \hspace{3.5cm} - ~
                        \mbox{Tr} \Big( \left[ A_\mu (0) \, , \,  A_\nu (2p) \right] \, t^r \Big) \, \mbox{Tr} \Big( \left[ A_\nu (p) \, , \,  A_\mu (-3p) \right] \, t^r \Big)  
                        \nonumber \\
                        & & \hspace{4cm} + ~
                        \mbox{Tr} \Big( \left[ A_\mu (0) \, , \,  A_\nu (-3p) \right] \, t^r \Big) \, \mbox{Tr} \Big( \left[ A_\mu (p) \, , \,  A_\nu (2p) \right] \, t^r \Big)  
                        \nonumber \\
                        & & \hspace{4.5cm} - ~
                        \mbox{Tr} \Big( \left[ A_\mu (0) \, , \,  A_\nu (-3p) \right] \, t^r \Big) \, \mbox{Tr} \Big( \left[ A_\nu (p) \, , \,  A_\mu (2p) \right] \, t^r \Big)  
                        \Bigg\}
   \ , 
 \label{FF:F0:0p2p3p}  \\
F^{(1)} (p^2) & = & \,  4 \,  ~ \sum_r  ~ \Bigg\{
                        \mbox{Tr} \Big( \left\{ A_\mu (0) \, , \,  A_\nu (p) \right\} \, t^r \Big) \, \mbox{Tr} \Big( \left\{ A_\mu (2p) \, , \,  A_\nu (-3p) \right\} \, t^r \Big)  
                        \nonumber \\
                        & & \hspace{2.5cm} + ~
                        \mbox{Tr} \Big( \left\{ A_\mu (0) \, , \,  A_\nu (p) \right\} \, t^r \Big) \, \mbox{Tr} \Big( \left\{ A_\nu (2p) \, , \,  A_\mu (-3p) \right\} \, t^r \Big)  
                        \nonumber \\
                        & & \hspace{3cm} + ~
                        \mbox{Tr} \Big( \left\{ A_\mu (0) \, , \,  A_\nu (2p) \right\} \, t^r \Big) \, \mbox{Tr} \Big( \left\{ A_\mu (p) \, , \,  A_\nu (-3p) \right\} \, t^r \Big)  
                        \nonumber \\
                        & & \hspace{3.5cm} + ~
                        \mbox{Tr} \Big( \left\{ A_\mu (0) \, , \,  A_\nu (2p) \right\} \, t^r \Big) \, \mbox{Tr} \Big( \left\{ A_\nu (p) \, , \,  A_\mu (-3p) \right\} \, t^r \Big)  
                        \nonumber \\
                        & & \hspace{4cm} + ~
                        \mbox{Tr} \Big( \left\{ A_\mu (0) \, , \,  A_\nu (-3p) \right\} \, t^r \Big) \, \mbox{Tr} \Big( \left\{ A_\mu (p) \, , \,  A_\nu (2p) \right\} \, t^r \Big)  
                        \nonumber \\
                        & & \hspace{4.5cm} + ~
                        \mbox{Tr} \Big( \left\{ A_\mu (0) \, , \,  A_\nu (-3p) \right\} \, t^r \Big) \, \mbox{Tr} \Big( \left\{ A_\nu (p) \, , \,  A_\mu (2p) \right\} \, t^r \Big)  
                        \Bigg\}
                         \ ,
 \label{FF:F1:0pp23p}  \\
F^{(2)} (p^2) & = & 4 ~ 
       \Bigg\{ 
                        \mbox{Tr} \Big( A_\mu (0) \, \,  A_\mu (p) \Big) \, \mbox{Tr} \Big( A_\nu (2p) \, \,  A_\nu (-3p)  \, \Big)
                        \nonumber \\
                         & & \hspace{1cm}
                         + ~   \mbox{Tr} \Big(  A_\mu (0) \,  \,  A_\mu (2p) \Big) \, \mbox{Tr} \Big( A_\nu (p) \,  \,  A_\nu (-3p) \Big) 
                        \nonumber \\
                         & & \hspace{1.5cm}
                         + ~   \mbox{Tr} \Big(  A_\mu (0) \,  \,  A_\mu (-3p) \Big) \, \mbox{Tr} \Big( A_\nu (p) \,  \,  A_\nu (2p) \Big) 
                        \nonumber \\
                         & & \hspace{2cm}
                         + ~   \mbox{Tr} \Big(  A_\mu (0) \,  \,  A_\nu (p) \Big) \, \mbox{Tr} \Big( A_\mu (2p) \,  \,  A_\nu (-3p) \Big) 
                        \nonumber \\
                         & & \hspace{2.5cm}
                         + ~   \mbox{Tr} \Big(  A_\mu (0) \,  \,  A_\nu (p) \Big) \, \mbox{Tr} \Big( A_\nu (2p) \,  \,  A_\mu (-3p) \Big) 
                        \nonumber \\
                         & & \hspace{3cm}
                         + ~   \mbox{Tr} \Big(  A_\mu (0) \,  \,  A_\nu (2p) \Big) \, \mbox{Tr} \Big( A_\mu (p) \,  \,  A_\nu (-3p) \Big) 
                        \nonumber \\
                         & & \hspace{3.5cm}
                         + ~   \mbox{Tr} \Big(  A_\mu (0) \,  \,  A_\nu (2p) \Big) \, \mbox{Tr} \Big( A_\nu (p) \,  \,  A_\mu (-3p) \Big) 
                        \nonumber \\
                         & & \hspace{4cm}
                         + ~   \mbox{Tr} \Big(  A_\mu (0) \,  \,  A_\nu (-3p) \Big) \, \mbox{Tr} \Big( A_\mu (p) \,  \,  A_\nu (2p) \Big) 
                        \nonumber \\
                         & & \hspace{4.5cm}
                         + ~   \mbox{Tr} \Big(  A_\mu (0) \,  \,  A_\nu (-3p) \Big) \, \mbox{Tr} \Big( A_\nu (p) \,  \,  A_\mu (2p) \Big) 
                         \Bigg\}
      \label{FF:F2:0p2p3p}
\end{eqnarray}
and for the (K. Conf. 3) kinematics 
\begin{center}
\begin{displaymath}
\boxed{ p_1 =  p_2 = p_3 = p,  ~ p_4 = - \, 3 \, p }
\end{displaymath}
\end{center}
\begin{eqnarray}
F^{(0)} (p^2) & =  &  \, 24 ~ \sum_r  ~ 
                        \mbox{Tr} \Big( \left[ A_\mu (p) \, , \,  A_\nu (p) \right] \, t^r \Big) \, \mbox{Tr} \Big( \left[ A_\nu (p) \, , \,  A_\mu (-3p) \right] \, t^r \Big)  \ ,
 \label{FF:F0:ppp3p}  \\
F^{(1)} (p^2) & = & 24 ~
       \sum_r  ~ 
                        \mbox{Tr} \Big( \left\{ A_\mu (p) \, , \,  A_\nu (p) \right\} \, t^r \Big) \, \mbox{Tr} \Big( \left\{ A_\mu (p) \, , \,  A_\nu (-3p) \right\} \, t^r \Big) 
                         \ ,
 \label{FF:F1:ppp3p}  \\
F^{(2)} (p^2) & = & 12 ~ 
       \Bigg\{ 
                        \mbox{Tr} \Big( A_\mu (p) \, \,  A_\mu (p) \Big) \, \mbox{Tr} \Big( A_\nu (p) \, \,  A_\nu (-3p)  \, \Big)
                        \nonumber \\
                         & & \hspace{2cm}
                         + \, 2 \,\,  \mbox{Tr} \Big(  A_\mu (p) \,  \,  A_\nu (p) \Big) \, \mbox{Tr} \Big( A_\mu (p) \,  \,  A_\nu (-3p) \Big) 
                         \Bigg\} \ .
      \label{FF:F2:ppp3p}
\end{eqnarray}
The expressions in Eqs (\ref{FF:F0:0pp2p}) to (\ref{FF:F2:ppp3p}) will be used in the simulations to measure
each of the lattice form factors $F^{(i)}$.

\bibliographystyle{apsrev4-1}
\bibliography{four_gluon_vertex_V4}

\end{document}